\documentclass[11pt, final]{article}
\usepackage{LTnotes}

\title{One-loop effect in the charged 2D black hole near extremality}
\author{Lorenzo Toni}

\affiliation{
	Center for Quantum Mathematics and Physics (QMAP)\\
	Department of Physics \& Astronomy, University of California, Davis, CA 95616 USA}

\emailAdd{ltoni@ucdavis.edu}

\abstract{We study the one-loop correction to the near-extremal quantum entropy of the charged two-dimensional black hole introduced in \cite{McGuigan:1992}. In target space this background can be understood as arising from the dimensional reduction of a three-dimensional solution of the low-energy string effective action. On the other hand, its worldsheet description is provided by the dimensionally reduced $\frac{SL(2, \mathbb{R}) \times U(1)}{U(1)}$ WZW model. Using this latter formulation, we compute the torus partition function to extract the temperature dependence of the one-loop stringy correction to the quantum entropy arising from the short-string sector. We show that this result is under control as long as the black hole temperature satisfies the condition $T \gtrsim \frac{1}{4 \pi \sqrt{k}}$, where the saturation of this bound corresponds to the long-string contribution to the free energy becoming divergent. We speculate that this threshold behavior of the partition function can be interpreted as a signature of the black hole/string transition.}

\begin{document} 
\maketitle


\section{Introduction}\label{sec:intro}

Two-dimensional black holes that admit an exact description in terms of worldsheet CFTs occupy a special place in string theory. Their solvability makes them powerful laboratories for addressing conceptual problems in gravity that are otherwise difficult to access. General relativity is well known to exhibit various pathologies, with spacetime singularities among the most severe. It is widely expected that a consistent theory of quantum gravity should provide insight into their resolution, and exact string-theoretic backgrounds offer a concrete setting in which such questions can be investigated.

Black hole solutions in string theory were first uncovered as classical solutions of the low-energy string effective action in ten spacetime dimensions \cite{HorowitzStrominger:1991}; see \cite{Horowitz:1992} for a comprehensive overview in four spacetime dimensions. While this approach is closely related to solving the beta-function equations of the worldsheet theory, it did not initially provide a direct construction of the underlying CFT. A major advance came with Witten’s realization \cite{Witten:1991} that a two-dimensional neutral black hole can be described exactly by a gauged WZW model \cite{Gawedzki:1988, Gawedzki:1989, Gawedzki:1991} based on the coset $\frac{SL(2, \mathbb{R})}{U(1)}$. Depending on whether a space-like or time-like $U(1)$ subgroup is gauged, one obtains Euclidean or Lorentzian black hole geometries: the two are, of course, related by analytic continuation. This construction made it possible to define an exact CFT governing string propagation in a curved black hole background. Geometrically, the resulting two-dimensional black hole resembles a radial slice of the four-dimensional Schwarzschild solution. Moreover, it was shown in \cite{Dijkgraaf:1992} that for such black holes physics beyond the singularity effectively decouples from observables accessible outside the horizon. Further insights into the $\frac{SL(2, \mathbb{R})}{U(1)}$ coset model and its connection to the Sine-Liouville theory \cite{Fukuda:2001, Giribet:2021}, along with a related dual matrix model, were explored in \cite{Kazakov:2002, Teschner:1999, Teschner:2000, Hikida:2009, Ribault:2005}.

A different picture emerges once gauge fields are included. In \cite{McGuigan:1992} the beta-function equations of a general worldsheet CFT with a gauge field were solved, yielding a charged two-dimensional black hole. The resulting geometry, which closely resembles the maximally extended Reissner–Nordstrom solution in four dimensions, can also be derived through dimensional reduction of the solution to the equations of motion of the low-energy string effective action in three dimensions. Subsequently, this two-dimensional charged black hole was reinterpreted as an exact coset CFT arising from the gauged WZW model $\frac{SL(2, \mathbb{R}) \times U(1)}{U(1)}$ in \cite{Giveon:1994, Giveon:2003}. In this approach, one begins with a four-dimensional $SL(2, \mathbb{R}) \times U(1)$ background, gauges an appropriate $U(1)$ subgroup to obtain a three-dimensional geometry, and then performs a dimensional reduction to arrive at a charged two-dimensional black hole. As in the neutral case, the choice of gauged subgroup determines the Lorentzian or Euclidean signature. An important qualitative difference from the uncharged case was uncovered in \cite{Giveon:2003}: scattering states originating beyond the singularity are no longer fully reflected. Consequently, initial data placed beyond the singularity can influence physical observables accessible to external observers. Furthermore, the thermodynamic properties of this background were investigated in \cite{Nappi:1992, Giveon:2004, Giveon:2005}. Related black hole constructions were explored in \cite{Mandal:1991, Ishibashi:1991, Horne:1992, Gibbons:1992, Johnson:1994, Antoniadis:1994}. 

In the present work, we focus on the class of charged two-dimensional black holes derived in \cite{McGuigan:1992, Giveon:2003} and compute the one-loop correction to their entropy. This analysis is particularly valuable for understanding the behavior of these objects in the near-extremal regime, where quantum effects are expected to dominate over classical contributions. 

The thermodynamic behavior of black holes in the near-extremal regime has been a central topic of investigation over the past two decades, as it provides a useful theoretical laboratory for exploring quantum effects in semiclassical gravity. The key observation is that the quantum entropy of a black hole develops a temperature-dependent logarithmic correction as one approaches the extremal limit. This behavior is closely tied to the structure of the near-horizon geometry: a near-extremal black hole contains an $AdS_2$ throat exhibiting an enhanced $SL(2,\mathbb{R})$ symmetry. This symmetry supports large diffeomorphism and large gauge transformation modes, which turn out to be exact zero modes \cite{Sen:2012} and play a crucial role in producing the logarithmic correction of the entropy \cite{Stanford:2017, Mertens:2023}. This phenomenon was first established for the BTZ black hole in \cite{Ghosh:2020}, both from a bulk perspective and via its dual boundary description. This analysis was later extended to the Reissner–Nordstrom black hole, using both a dimensional reduction of the near-horizon region to JT gravity \cite{Nayak:2018, Iliesiu:2020, Iliesiu:2021} and a gravitational path integral approach \cite{Iliesiu:2022}. More recently, similar considerations have been applied to the Kerr black hole in four-dimensional asymptotically flat spacetime \cite{Rakic:2024, Kapec:2024}, Kerr and Kerr-Newman black holes in Anti-de-Sitter spacetime \cite{Maulik:2026}, and the Reissner-Nordstrom black hole in de-Sitter spacetime \cite{Maulik2:2026}. The case of the BTZ black hole was elaborated further from a string-theoretic viewpoint in \cite{Ferko:2025}. It is important to emphasize that this discussion applies only to non-supersymmetric black holes; see \cite{Banerjee:2011, Banerjee1:2011, Sen:2012, Heydeman:2022, Iliesiu:2025} for an overview of the supersymmetric case.

Since our charged two-dimensional black hole reduces to the usual $AdS_2$ geometry with background gauge fields in the extremal limit \cite{Giveon:2005}, we are able to reproduce the semiclassical logarithmic correction to the near-extremal entropy by adapting the analysis of \cite{Iliesiu:2021}.
However, because this black hole admits a microscopic string-theoretic description, the semiclassical result can be refined by incorporating genuine stringy effects. Our primary goal is then to compute the one-loop correction to the entropy using an entirely string-theoretic framework. Our general strategy closely follows the analysis of \cite{Maldacena:2001}, where the one-loop free energy of strings propagating on $AdS_3$ was derived from the torus partition function, as well as the approach taken in \cite{Hanany:2002}, where the authors determine the spectrum of the uncharged two-dimensional black hole using path-integral methods. More explicitly, we evaluate the torus partition function of the $\frac{SL(2, \mathbb{R}) \times U(1)}{U(1)}$ WZW model in the canonical ensemble and, after dimensional reduction along the compact $U(1)$ direction, we extract the temperature dependence of the stringy correction to the black hole quantum entropy. 

Given our primary interest lies in the low-temperature limit of the entropy, it is natural to focus on the contribution arising from short-string states, which correspond to discrete representations of the $\mathfrak{sl}_k(2, \mathbb{R})$ current algebra \cite{Maldacena1:2001}. In contrast, long-string states, associated with continuous $\mathfrak{sl}_k(2, \mathbb{R})$ representations, dominate the entropy at high temperature. Despite the restriction to the short-string sector, we are unable to derive a closed-form expression for the full one-loop correction. The main technical obstruction is the path integral over the gauge-field holonomy, which also prevents us from recovering the full black hole spectrum in the present work. We defer this goal to a future publication, where we also plan to provide a more systematic treatment of the long-string contribution.

The one-loop correction we obtain from the short-string sector is shown to be valid provided that the black hole temperature satisfies $T \gtrsim \frac{1}{4 \pi \sqrt{k}}$. Although this condition restricts our result to hold strictly above extremality, sufficiently large values of $k$ allow us to probe the near-extremal regime while keeping perturbative control. Since approaching this lower bound causes the long-string contribution to the free energy to diverge, we speculate that a natural physical interpretation for this threshold behavior is in terms of the black hole/string transition. 

To understand how such a transition can occur in the low-temperature limit, it is instructive to recall the mechanism established in the context of the uncharged 2D black hole \cite{Horowitz:1997, Giveon_transition:2005, Kutasov:2005, Giveon:2007}. In that setting, increasing the black hole temperature triggers severe thermal fluctuations near the horizon. Upon reaching the Hagedorn temperature, the description based on the canonical ensemble breaks down, signaling the failure of conventional equilibrium thermodynamics \cite{Atick:1988}. In this regime, injected energy no longer increases the temperature but instead feeds the exponential growth of string states. Consequently, the classical black hole description ceases to be valid and must be replaced by a string-theoretic one in terms of long, highly excited fundamental strings. In the Euclidean geometry, this transition is mediated by tachyonic winding modes wrapping the thermal circle. These modes become effectively massless at the transition and dominate the thermodynamics. The resulting phase is therefore characterized by a winding condensate, signaling a reorganization of the underlying degrees of freedom \cite{Berkooz:2007, Brustein:2021}.

An analogous transition for the charged 2D black holes was explored in \cite{Giveon:2006}. Crucially, the string/black hole transition for charged black holes can occur at temperatures that are arbitrarily low compared to the Hagedorn temperature. In particular, both the black hole and Hagedorn temperatures vanish simultaneously in the extremal limit. This interplay between the Hagedorn transition and the low-temperature limit of this charged black hole may mirror the mechanism discussed in \cite{Chen:2021}, where the extremal limit of the charged Horowitz-Polchinski string \cite{Horowitz:1998} is shown to map precisely to the Hagedorn transition of the corresponding neutral solution.

The paper is organized as follows. In Section \ref{sec:extremal}, we review the semiclassical analysis of the one-loop correction to the entropy of the four-dimensional Reissner-Nordstrom black hole in asymptotically flat spacetime from a path integral perspective \cite{Iliesiu:2022}. This discussion provides the benchmark for the analysis of the charged two-dimensional black hole. In particular, Section \ref{sec:target_analysis} discusses its target-space origin, obtained via dimensional reduction of a three-dimensional solution of the low-energy string effective action, and analyzes its classical and semiclassical thermodynamic properties. In Section \ref{sec:ws_analysis} we present the complementary worldsheet description, in which the black hole is realized as a CFT arising from dimensional reduction of the $\frac{SL(2, \mathbb{R}) \times U(1)}{U(1)}$ WZW model. In Section \ref{sec:partition_fct}, we compute the torus partition function of the $\frac{SL(2, \mathbb{R}) \times U(1)}{U(1)}$ WZW model; technical details of the path-integral derivation are deferred to Appendices \ref{sec:path_int_freefields} and \ref{sec:path_int_ads3}. Modular invariance of the resulting expression is explicitly verified in Section \ref{sec:modular_inv}. Finally, in Section \ref{sec:free_energy} we extract the temperature dependence of the leading one-loop correction to the black hole entropy associated with the short-string sector by dimensionally reducing the torus partition function and integrating over the torus moduli space. The details of the underlying computation are presented in Appendix \ref{sec:euclidean}. We conclude by discussing the structure of this one-loop correction and its connection to the black hole/string transition.

\section{Review of charged 4D black holes near extremality}\label{sec:extremal}

We begin by reviewing the semiclassical correction to the thermodynamics of near-extremal charged black holes, as this framework will play a central role in the interpretation of our results throughout the remainder of the paper. These geometries develop an $AdS_2\times \mathbb{S}^2$ throat near the horizon, which renders the semiclassical evaluation of the gravitational path integral tractable. For concreteness, we focus on the four-dimensional Reissner-Nordstrom black hole in asymptotically flat spacetime; however, the analysis could be straightforwardly generalized to include a non-vanishing cosmological constant.

In Euclidean signature, the Einstein-Maxwell action reads 
\begin{equation}\label{eq:em_action}
    I = \int_{\mathcal{M}} d^4 x \sqrt{g} \left( \frac{1}{16 \pi G_N} R - \frac{1}{4} F^2 \right) + \frac{1}{8 \pi G_N} \int_{\mathcal{\partial M}} d^3 x \sqrt{\gamma} K , 
\end{equation}
where $R$ and $F$ represent the usual Ricci scalar and the gauge field strength, respectively, in four dimensions. The Gibbons-Hawking-York boundary term ensures a well-defined variational principle. The relevant black hole solution is (in the convention $G_N=1$)
\begin{equation}
    \begin{split}
        ds^2 & = f(r) \, d \tau^2 + \frac{dr^2}{f(r)} + r^2 \left( d \theta^2 + \sin^2(\theta) d \phi^2 \right), \qquad f(r) = 1-\frac{2 \, M}{r} + \frac{Q^2}{r^2}, \\
        A & = - i \, Q \left( \frac{1}{r_+} - \frac{1}{r} \right) \, d\tau,
    \end{split}
\end{equation}
where $M$ and $Q$ are the mass and the charge of the black hole, respectively, and $r_\pm = M \pm \sqrt{M^2 - Q^2}$ are the inner and outer horizons. Applying the standard semiclassical analysis, one finds that the Hawking temperature and the chemical potential are
\begin{equation}\label{eq:temp_chem_rn_4d}
    T = \frac{r_+ - r_-}{4 \,\pi \,r_+^2}, \qquad \mu= \frac{Q}{r_+}.
\end{equation}
The classical entropy is given by the Bekenstein-Hawking formula $S_0 = \pi r_+^2$. In the extremal limit $M = Q$, the temperature vanishes and the entropy reduces to $\pi \, Q^2$. Defining the radial deviation from the outer horizon by $\rho= r - r_+$, the near-horizon region of the extremal black hole, to leading order in $\rho \ll r_+$, reduces to $AdS_2 \times \mathbb{S}^2$:
\begin{equation}\label{eq:zero_order}
    \begin{split}
        ds_{(0)}^2 & = Q^2 ( \sinh^2(\rho) d \tau^2 + d \rho^2), \\
        A_{(0)}& = i Q (\cosh(\rho) -1) d \tau.
    \end{split} 
\end{equation}

Away from extremality, the near-horizon region no longer factorizes exactly as $AdS_2 \times \mathbb{S}^2$, but instead forms a finite throat that interpolates between the asymptotically flat region and the horizon. To make this explicit, we work in the canonical ensemble\footnote{In the canonical ensemble a boundary term for the field strength has to be added to the action $(\ref{eq:em_action})$ in order for the variational principle to be well-defined \cite{Iliesiu:2022}.} where the charge $Q$ is held fixed. In this ensemble, the mass can be expanded as a power series in the temperature \cite{Iliesiu:2022}
\begin{equation}
    M= Q + 4 \, \pi^2 \, Q^3 \, T^2 + 16 \, \pi^3 \, Q^4 \, T^3 + \mathcal{O}(T^4),
\end{equation}
that in turn leads to a temperature expansion for $r_+(T)$. Considering the change of coordinates
\begin{equation}
    r \rightarrow r_+(T) + 2 \, \pi \, Q \, T \left(\cosh(\rho) -1 \right),
\end{equation}
the first order correction to the black hole solution near extremality is
\begin{equation}\label{eq:first_order_rn}
    \begin{split}
        ds_{(1)}^2  = & \,\pi Q^3 T (\cosh(\rho) +2)\tanh^2 \left( \frac{\rho}{2} \right) ( - \sinh^2(\rho) d \tau^2 + d \rho^2) \\
        & + 4 \pi Q^3 T \cosh(\rho) (d \theta^2 + \sin^2(\theta) d \phi^2), \\
       A_{(1)} = & \, - 2 \pi i Q^2 T \sinh^2(\rho)  d \tau.
    \end{split}
\end{equation}

We are now in position to discuss the quantum corrections to the near-extremal entropy. Within the semiclassical approximation, the partition function is evaluated by expanding the corresponding gravitational path integral around a classical saddle point. Since the near-horizon geometry of the near-extremal black hole, defined by ($\ref{eq:zero_order}$) and $(\ref{eq:first_order_rn})$, is an exact solution of the Einstein-Maxwell equations order by order in $T$, it defines a legitimate saddle around which the path integral may be evaluated. More explicitly,
\begin{equation}\label{eq:formula_target_part}
    Z = \int \mathcal{D}h \, \mathcal{D} \mathcal{A} \, e^{- I \left( g_{(0)}+g_{(1)}, \, A_{(0)}+A_{(1)} \right) - I_{\text{quadr}}(h,\mathcal{A}) \, +\, ...} =  \exp \left( S_0 + S_1 + \, ... \,\right),
\end{equation}
where $h$ and $\mathcal{A}$ denote the quantum fluctuations of the metric and gauge field around the classical background, subject to boundary conditions appropriate for the canonical ensemble. Expanding the action to quadratic order captures the one-loop effects of these fluctuations. Consequently, the exponent in the right-hand side defines a perturbative expansion for the so called quantum entropy \cite{Banerjee:2011, Banerjee1:2011, Sen:2012}: the on-shell value of the action $S_0$ corresponds to the classical near-extremal value, whereas $S_1$ represents the leading one-loop quantum correction.

The evaluation of the one-loop contribution $S_1$ is performed using heat-kernel methods, which systematically account for the spectrum of quadratic fluctuations. A crucial aspect of this analysis is the proper handling of zero modes, which are associated with large diffeomorphisms and large gauge transformations. As shown in \cite{Iliesiu:2022}, the only temperature-dependent correction to the entropy arises from metric zero modes corresponding to large diffeomorphisms on $AdS_2$. These modes give rise to the universal Schwarzian sector \cite{Stanford:2017, Mertens:2023, Ghosh:2020} that emerges upon dimensional reduction of the near-horizon geometry to two dimensions \cite{Iliesiu:2021}, where the effective JT gravity description becomes applicable. 

This conclusion is even more apparent from the discussion in \cite{Kolanowski:2025}. Indeed, the eigenvalues of the Schwarzian modes are proportional to $T/T_q^{\,\text{Schw}}$, where $T_q^{\text{Schw}}$ sets the temperature scale at which these modes become strongly coupled. Since this scale can be identified from classical thermodynamics through the relation \cite{Sachdev:2019}
\begin{equation}\label{eq:schw_coupl}
    T_q^{\, \text{Schw}} = 4 \, \pi^2 \left( \frac{\partial S}{\partial T}\Big|_{T=0} \right)^{-1},
\end{equation}
the Schwarzian coupling is determined by the linear-in-temperature term in the classical near-extremal entropy. By analogy, the corresponding temperature scale for the gauge modes is controlled by the inverse charge susceptibility at extremality:
\begin{equation}\label{eq:charge_susc}
    T_q^{\, U(1)} = \left( \frac{\partial Q}{\partial \mu} \Big|_{T=0} \right)^{-1}.
\end{equation}
From (\ref{eq:temp_chem_rn_4d}) we observe that the chemical potential approaches a constant value, independent of the charge, in the extremal limit. Consequently, the charge susceptibility diverges at extremality, implying that the associated coupling for the gauge modes vanishes.

Overall, including only the temperature-dependent corrections, the near-extremal quantum entropy in the canonical ensemble takes the following form:
\begin{equation} \label{eq:rn_entropy}
    \ln(Z) \, = \, \pi  Q^2  + 4  \pi^2 Q^3  T +  \frac{3}{2}  \ln(Q^3 T).
\end{equation}
The first two terms represent the classical contributions: the first gives the entropy at extremality, while the second accounts for the classical correction due to small deviations away from it. In contrast, the logarithmic term stems from genuinely quantum effects, which dominates the infrared regime. The quantum entropy becomes singular when $T = 0$, indicating that truly extremal Reissner-Nordstrom black holes do not exist in perturbative semiclassical gravity.

The remainder of this paper is devoted to deriving the one-loop correction to the quantum entropy of the charged 2D black hole. This geometry was originally introduced in \cite{McGuigan:1992} and later understood as the dimensional reduction of the $\frac{SL(2, \mathbb{R}) \times U(1)}{U(1)}$ WZW model \cite{Giveon:2003}. In Section \ref{sec:target_analysis} we show that a semiclassical target-space analysis yields an expansion for the quantum entropy analogous to \eqref{eq:rn_entropy}, whereas in Section \ref{sec:quantum} we refine this semiclassical result by incorporating genuine stringy effects.

\section{Charged 2D black hole}\label{sec:initial}

In this section, we introduce the class of charged 2D black holes under consideration. On one hand, these backgrounds originate from the dimensional reduction of a broader family of three-dimensional solutions of the target-space string effective action. On the other, we analyze their string-theoretic origin in terms of the corresponding worldsheet description, formulated explicitly through the $\frac{SL(2, \mathbb{R}) \times U(1)}{U(1)}$ WZW model.

\subsection{Target space analysis}\label{sec:target_analysis}

The relevant low-energy string effective action in three spacetime dimensions is \cite{Horowitz:1993}
\begin{equation}
    I_{(3)} = \int d^3 x \, \sqrt{-g} \,  e^{-2\Phi} \left( R  + 4 \left( \nabla \Phi \right)^2 - \frac{1}{12} H^2  + \frac{4}{k} \right),
\end{equation}
where $R$ is the Ricci scalar of the background spacetime, $H$ the field strength of the corresponding two-form $B$-field and $\Phi$ the dilaton field. The last term represents a negative cosmological constant contribution, the constant $k$ being associated with the level of the underlying current algebra. A family of solutions of the equations of motion of the above action is given by
\begin{equation}\label{eq:3dmetric}
    \begin{split}
        ds_{(3)}^2 & =  - \frac{k}{\coth^2(\rho) -a^2} d t^2 + k d \rho^2 + \frac{1}{2} d x^2 + \frac{\sqrt{2k} \, a}{\coth^2(\rho)-a^2} dt \, dx, \\
        B &=\frac{\sqrt{2k} \, a}{\coth^2(\rho)-a^2} \ d x\wedge  d t, \\
        \Phi&=\Phi_0-\frac{1}{2}\ln \left( \cosh^2(\rho)-a^2\sinh^2(\rho) \right),
    \end{split}
\end{equation}
where $\rho \in [0, \infty)$, $t \in \mathbb{R}$ and $x \in \mathbb{R}$. Moreover, $a$ is an arbitrary constant and $\Phi_0$ is the asymptotic value of the dilaton. We will show in Section $\ref{sec:ws_analysis}$ that the constant $a$ is actually related to the coupling constants of the underlying worldsheet CFT. The above classical solution represents the dominant background valid in the limit $k \gg 1$, where quantum fluctuations are suppressed. Geometrically, it represents a 3D boosted black string belonging to the same family of black string solutions introduced in \cite{Horne:1992}. 

Compactifying the $x$-direction on a circle of radius $ R \ll \sqrt{k}$, the dimensional reduction of the solution ($\ref{eq:3dmetric}$) yields \cite{Chen:2022}
\begin{equation} \label{eq:2Dbh}
    \begin{aligned}
        ds_{(2)}^2 &= k \left[-\left(\frac{\coth(\rho)}{\coth^2(\rho)-a^2}\right)^2dt^2 + d \rho^2\right], \\
        A &=\frac{\sqrt{2k} \, a}{\coth^2(\rho)-a^2}dt, \\
        \Phi&=\Phi_0-\frac{1}{2}\ln \left( \cosh^2(\rho)-a^2\sinh^2(\rho) \right).
    \end{aligned}
\end{equation}
It is well known that this background represents a charged 2D black hole \cite{Dijkgraaf:1992, Giveon:2003}, where the gauge field $A$ is associated with the translation charge in the $x$-direction. Moreover, upon dimensional reduction of the $B$-field we get another gauge field
\begin{equation}\label{eq:second_field}
    A' = \frac{\sqrt{2k} \, a}{2(\coth^2(\rho)-a^2)}dt,
\end{equation}
which measures the winding charge in the $x$-direction. Therefore, this 2D black hole carries both momentum and winding charge around the compactified circle. 

It is worth mentioning that ($\ref{eq:2Dbh}$) may be also obtained independently as classical solution of the  low-energy string effective action in two spacetime dimensions
\begin{equation}
    I_{(2)} = \int d^2 x \, \sqrt{-g} \, e^{-2 \Phi} \left( R  + 4 \left( \nabla \Phi \right)^2 - \frac{1}{8} F^2 - \frac{1}{2} F'\,^2  + \frac{4}{k} \right).
\end{equation}
The equations of motion of the above action represent the beta functions for a worldsheet CFT which naturally emerges from an appropriate heterotic string compactification, as originally discussed in \cite{McGuigan:1992}. However, as we will extensively discuss in the next section, the black hole under consideration also possesses a purely worldsheet interpretation in terms of the $\frac{SL(2, \mathbb{R}) \times U(1)}{U(1)}$ WZW model.  

Regardless of its origin, the black hole in ($\ref{eq:2Dbh}$) admits an interpretation as the 2D analogue of the 4D Reissner-Nordstrom black hole \cite{Giveon:2003, McGuigan:1992}. Indeed, upon the change of coordinates 
\begin{equation} \label{eq:change_coord}
    v \, = \, \frac{2 \, t}{1-a^2}, \qquad r \, = \, \frac{2 \, e^{-2 \Phi}}{\sqrt{k}},
\end{equation}
the black hole ($\ref{eq:2Dbh}$) takes the form 
\begin{equation}\label{eq:2d_rn}
    \begin{split}
        d s^2_{(2)} &=  \frac{k}{4} \left( - f(r) \, dv^2 + \frac{d r^2}{r^2 \, f(r)} \right), \qquad f(r) \, = \, 1 - \frac{2 \, M}{r} + \frac{Q^2}{r^2}, \\
         A  &  =  \sqrt{\frac{k}{2}} \, Q \, \left( \frac{1}{r_+} - \frac{1}{r} \right) \, dv, \\
         \Phi & = - \frac{1}{2} \, \ln \left( \frac{\sqrt{k} \, r}{2} \right).
    \end{split}
\end{equation}
The corresponding causal structure has been analyzed in detail in \cite{Giveon:2003}. The singularity is located at $r=0$ and the inner and outer horizons are
\begin{equation}\label{eq:horizon}
    r_\pm \, = M \, \pm \, \sqrt{M^2 - Q^2} \, \Longrightarrow \,
    \begin{dcases}
        r_+ = \frac{Q}{a}, \\
        r_- = Q\,a,
    \end{dcases}
\end{equation}
whereas the ADM mass and charge are given by
\begin{equation}\label{eq:mass}
    M= \frac{e^{-2 \Phi_0}}{\sqrt{k}} \, (1 + a^2), \qquad Q=  \frac{2 \, a \, e^{-2 \Phi_0}}{\sqrt{k}}.
\end{equation}
Applying the standard semiclassical analysis, one finds that the Hawking temperature and the chemical potential of the black hole are \cite{Giveon:2004}
\begin{equation} \label{eq:classical_temp}
    T= \frac{1-a^2}{2 \pi \sqrt{k}}, \qquad \mu = \sqrt{\frac{k}{2}} \, \frac{Q}{r_+},
\end{equation}
while the classical entropy and heat capacity (at constant charge) take the form
\begin{equation}\label{eq:classical_entropy}
    \begin{split}
        S_0  & = \pi \, \sqrt{k} \, r_+ = \frac{\pi \sqrt{k} \, Q}{\sqrt{1-2 \pi \sqrt{k} \, T}}, \\
        C_Q  & = \frac{\pi \sqrt{k} \, (r_+ - r_-)}{2} \, \frac{r_+}{r_-} = \frac{\pi^2 \, k \, Q \, T}{\left(1 - 2 \pi \sqrt{k}\, T \right)^{3/2}}.
    \end{split}
\end{equation} 

Naively, at extremality $a = 1$ one has $M=Q$ and the horizon ($\ref{eq:horizon}$) becomes degenerate. As a consequence, the temperature and the heat capacity vanish, whereas the entropy and the chemical potential approach finite values $\pi \sqrt{k}\,Q$ and $\sqrt{\frac{k}{2}}$, respectively. Defining the deviation from the horizon as $u = r-Q$ and retaining only the terms with $u \ll Q$, the near-horizon geometry reduces to that of $AdS_2$ with a background gauge field \cite{Giveon:2005}
\begin{equation} \label{eq:metric_extremal}
    \begin{split}
        d s^2_{(2)} & = \frac{k}{4}  \left( - u^2 \, dv^2 + \frac{d u^2}{u^2} \right),\\
        A & = \sqrt{\frac{k}{2}} \, u \, dv, \\
        \Phi & = - \frac{1}{2} \ln \left( \frac{\sqrt{k} \, Q}{2} \right),
     \end{split}
\end{equation}
where the coordinates have been appropriately rescaled.

However, the coordinate transformation $(\ref{eq:change_coord}$) becomes ill-defined in the limit $a \to 1$, so only the near-extremal limit can be reliably analyzed. In the canonical ensemble, this limit has to be taken while keeping the charge $Q$ fixed, that is, with $a / \sqrt{k}=\,$constant. To study this regime we follow the strategy outlined in Section \ref{sec:extremal}, namely we express the mass and the outer horizon as a power series in the temperature ($\ref{eq:classical_temp}$) while holding the charge fixed: 
\begin{equation}
    \begin{split}
        M & = Q+  \frac{1}{2} \pi^2 k \, Q \, T^2  + \mathcal{O}(T^3), \\
        r_+ & = Q + \pi  \sqrt{k} \, Q \,T + \mathcal{O}(T^2).
    \end{split}
\end{equation}
By retaining only the first order correction in the temperature, the geometry ($\ref{eq:2d_rn}$) in the near-horizon near-extremal limit becomes
\begin{equation} \label{eq:near_ext_saddle}
    \begin{split}
        ds_{(2)}^2 & = \frac{k}{4} \left( - \left(u^2 - \pi^2 k \, Q^2 T^2\right) d v^2 + \frac{d u^2}{u^2 -\pi^2 k \, Q^2 T^2} \right), \\
        A & = \sqrt{\frac{k}{2}} \left( u - \pi \sqrt{k}\, Q\, T \right) dv, \\
        \Phi & = - \frac{1}{2} \ln \left( \frac{\sqrt{k} \, Q}{2} \right).
    \end{split}
\end{equation}
This background represents a finite-temperature deformation of the infinite $AdS_2$ throat characteristic of the extremal geometry ($\ref{eq:metric_extremal})$, with the deviation proportional to the temperature. The corresponding entropy and heat capacity follow from a low-temperature expansion of $(\ref{eq:classical_entropy})$: 
\begin{equation}\label{eq:near_ext_entropy}
    \begin{split}
        S_0 & = \pi  \sqrt{k} \, Q + \pi^2 k \, Q \, T + \mathcal{O}(T^2), \\
        C_Q & = \pi^2 k \, Q \, T + 3 \pi ^3 k^{3/2} Q \, T^2+ \mathcal{O}(T^3).
    \end{split}
\end{equation}

At this stage one could treat the near-extremal geometry (\ref{eq:near_ext_saddle}) as a classical saddle point and compute quantum corrections to the near-extremal entropy (\ref{eq:near_ext_entropy}) by evaluating the gravitational path integral around it. However, this discussion can be carried out more directly by adapting the semiclassical analysis of the four-dimensional Reissner–Nordstrom
black hole in asymptotically Anti-de Sitter spacetime detailed in \cite{Iliesiu:2021}, where a complementary perspective to the gravitational path integral is employed. Rather than performing the path integral around the nearly $AdS_2 \times \mathbb{S}^2$ geometry arising in the near-horizon near-extremal regime, the authors first carry out a dimensional reduction on $\mathbb{S}^2$. This leads to the emergence of the effective JT gravity description on $AdS_2$, whose one-loop partition function \cite{Stanford:2017} allows one to recover the quantum corrections to the near-extremal black hole entropy.

It is important to emphasize a key distinction between the standard four-dimensional Reissner–Nordstrom black hole and the 2D model presented here. Specifically, our model incorporates two gauge fields, namely ($\ref{eq:2Dbh}$) and ($\ref{eq:second_field}$). In principle, both fields should influence the temperature dependence of the one-loop correction. However, as discussed around (\ref{eq:charge_susc}), the coupling for the gauge modes is determined by the inverse charge susceptibility at extremality. Since (\ref{eq:classical_temp}) shows that this quantity vanishes for both fields, the zero modes associated with their large gauge transformations do not contribute to the temperature dependence of the one-loop correction to the near-extremal quantum entropy.

Overall, by mapping the notation in \cite{Iliesiu:2021} to ours, the near-extremal quantum entropy for the black hole (\ref{eq:near_ext_saddle}) looks
\begin{equation}\label{eq:semi_2d_bh_entr}
    \ln(Z_{2D})\Big|_{\text{semiclassical}} = \pi  \sqrt{k} \, Q + \pi^2 k \, Q \, T + \frac{3}{2} \ln \left( \frac{k \, Q \, T}{4} \right),
\end{equation}
where the first two terms reproduce the classical result (\ref{eq:near_ext_entropy}), while the last term represents the standard logarithmic correction. This expression is analogous to the corresponding result for the Reissner–Nordstrom black hole in asymptotically flat spacetime ($\ref{eq:near_ext_entropy}$) and matches the expected qualitative behavior.

Given that the black hole ($\ref{eq:near_ext_saddle}$) also admits a microscopic string-theoretic description, this semiclassical result can be refined by incorporating genuine stringy effects. In what follows, we compute the torus partition function of the underlying worldsheet theory and extract from it the corresponding one-loop correction to the quantum entropy. The near-extremal behavior is then obtained by taking the appropriate low-temperature limit.

\subsection{Worldsheet analysis}\label{sec:ws_analysis}

Having introduced the target space interpretation of the classical charged 2D black hole in the previous section, we now turn to discuss its microscopic origin within string theory. From the worldsheet perspective, this background is naturally described in terms of a gauged WZW model. More specifically, the starting point is the WZW model \cite{Gawedzki:1988, Gawedzki:1989} based on the four-dimensional group manifold $SL(2, \mathbb{R}) \times U(1)$. The $U(1)$ component represents a decoupled compact boson, whereas the $SL(2, \mathbb{R})$ WZW model is commonly employed to describe strings propagating on $AdS_3$ \cite{Maldacena:2001}. In this setting, one generally takes the time coordinate to be non-compact, effectively working with the universal covering space of $SL(2, \mathbb{R})$. Since we will be working in Euclidean signature, the relevant group is actually $\frac{SL(2, \mathbb{C})}{SU(2)}$. The essential step in constructing the Euclidean black hole geometry is the gauging of an appropriate time-like $U(1)$ subgroup of $SL(2, \mathbb{R})$. This gauging procedure yields a three-dimensional spacetime with nontrivial metric and dilaton profiles \cite{Giveon:2003}. Upon dimensional reduction along the compact $U(1)$ direction, one arrives at the Euclidean version of the charged 2D black hole geometry given in $(\ref{eq:2Dbh})$.

More concretely, the Euclidean worldsheet action of the $SL(2, \mathbb{R}) \times U(1)$ WZW model is given by
\be
    I = \frac{k}{2 \pi} \int d^2 \sigma \Tr \left( g^{-1} \, \p_z g \, g^{-1} \, \p_{\bar{z}} g \right)+ \frac{i k}{12 \pi} \int_B \text{Tr} \left( g^{-1} d g \right)^3 + \frac{1}{\pi} \int d^2 \sigma \, \p_z x \, \p_{\bar{z}} x,
\ee
with $(g, x) \in SL(2, \mathbb{R}) \times U(1)$ such that $x \sim x + 2 \pi R$. Here we are omitting the decoupled worldsheet theory of the internal CFT, which is required for internal consistency. Since $H^3(SL(2,\mathbb{R}), \mathbb{R})=0$, the above action is independent of the choice of $B$ for any real value of $k$. At the algebraic level, the decoupled boson has central charge $c_{\mathfrak{u}(1)} =1$, whereas the $SL(2, \mathbb{R})$ realizes an affine $\mathfrak{sl}_k(2, \mathbb{R})$ current algebra at level $k$ with central charge $c_{\mathfrak{sl}_k(2, \mathbb{R})} = \frac{3 k}{k-2}$. Throughout this paper we will work in the regime $k > 2$.

To get an explicit expression for the action, we make the following choice of coordinates for the $SL(2, \mathbb{R})$ sector
\be
    g = 
    \begin{pmatrix}
        v & -e^\phi (1 + |v|^2) \\
        e^{-\phi} & - \bar{v}
    \end{pmatrix},
\ee
leading to
\be
    I = \frac{k}{2{\pi}} \int 2 \, d^2\sigma \left[\p_z \phi \, \p_{\bar z} \phi + (\p_z+\p_z \phi) \, \bar v \, (\p_{\bar z}+\p_{\bar z} \phi) \,v + \frac{1}{k} \, \p_{z} x \, \p_{\bar z} x \right].
\ee
One can clearly see that the target space geometry is that of Euclidean $AdS_3 \times \mathbb{S}^1$. Our objective is to analyze the target-space background obtained by gauging a timelike $U(1)$ subgroup of the isometry group. The action of this $U(1)$ component on the fields is specified by
\be
    \begin{split}
        v & \to v \, e^{\, \eta \, (p'-p)}, \quad \bar{v} \to \bar{v} \, e^{\, \eta \, (\bar{p}'-\bar{p})},\\
        \phi_R & \to \phi_R -p \, \eta, \quad \phi_L  \to \phi_L -\bar{p} \, \eta, \\
        x_R & \to x_R-P \, \eta, \quad x_L \to x_L-\bar{P} \, \eta ,
    \end{split}
\ee
where $\eta$ is the gauge parameter and $p$, $\bar{p}$, $p'$, $\bar{p}'$, $P$, $\bar{P}$ are constant coefficients characterizing the embedding of the gauge group into the global symmetry group. The gauge invariant action describing the resulting $\frac{SL(2, \mathbb{R}) \times U(1)}{U(1)}$ coset takes the form
\be\label{eq:action_gauge_wzw}
\begin{aligned}
    I = & \frac{k}{2{\pi}} \int 2 \, d^2\sigma \Bigg[ (\partial_z \phi+p \, A_z)(\partial_{\bar z} \phi+\bar{p} \, A_{\bar z})  \\
    & + (\partial_z+\partial_z \phi+ p' \, A_z) \, \bar v \, (\partial_{\bar z}+\partial_{\bar z} \phi+\bar{p}' \, A_{\bar z}) \, v  + \frac{1}{k} (\partial_{z} x+P \, A_{z}) (\partial_{\bar z} x+ \bar{P} \, A_{\bar z}) \Bigg],
\end{aligned}
\ee
where the worldsheet gauge fields transform as 
\be
    A_z\to A_z+\partial_z \eta, \quad  A_{\bar{z}}\to A_{\bar{z}}+\partial_{\bar{z}} \eta.
\ee
The central charge of this gauged WZW model is simply $c_{\text{wzw}}=\frac{3 k}{k-2}$ because the gauging procedure lowers the central charge by one. Hence, anomaly cancellation requires the internal CFT to have a central charge $c_{\text{int}}= 26 - \frac{3 k}{k-2}$.

For the following analysis, it turns out to be convenient to work with the global coordinates 
\be
    v = e^{i \theta} \, \sinh (\rho), \quad \bar{v} = e^{- i \theta} \, \sinh (\rho), \quad
    \phi = \xi- \ln \left( \cosh (\rho) \right).
\ee
Naively integrating out the gauge fields in the full action leads to an expression that obscures the underlying structure of the theory. To obtain a more transparent description, we restrict attention to the particular charge configuration $p=p'$, $\bar{p}=\bar{p}'$. Then, the coordinates $\rho$ and $\theta$ are manifestly gauge invariant, whereas the other fields transform non-trivially under a gauge transformation as
\be
    \begin{split}
        \p_{ z}\xi & \to \p_{ z}\xi-p \, \p_{ z}\eta, \quad \p_{ \bar{z}}\xi \to \p_{ \bar{z}}\xi-\bar{p} \, \p_{ \bar{z}}\eta, \\
        \quad  \p_{\bar z} x & \to \p_{\bar z} x-\bar{P} \, \p_{\bar z}  \eta, \quad  \p_{ z} x \to \p_{ z} x-P \, \p_{z} \eta.
    \end{split}
\ee
It is therefore natural to work with the following gauge invariant combinations:
\be
p \, \p_{ z} x-P \, \p_{ z} \xi, \quad \bar{p} \, \p_{ \bar{z}} x-\bar{P} \, \p_{ \bar{z}} \xi.
\ee 
As a result, the gauge invariant action takes the form
\be
    \begin{split}
        I = \frac{k}{2 \pi} & \int 2 \, d^2 \sigma \Bigg[ \p_z \rho \, \p_{\bar{z}} \rho +\p_z \theta \, \p_{\bar{z}} \theta \sinh^2 (\rho) \,  \left( 1- \frac{p \, \bar{p} \, \sinh^2 (\rho)}{p\, \bar{p}\, \cosh^2(\rho)+\frac{P \,\bar{P}}{k}} \right) \\
        & -i \, \tanh (\rho) \,\p_z \theta \, \p_{\bar{z}} \rho + i \, \tanh (\rho) \, \p_z \rho \, \p_{\bar{z}} \theta \\
        & + \frac{\cosh^2 (\rho)}{k \, (p \,\bar{p} \, \cosh^2(\rho)+\frac{P \, \bar{P}}{k})} \, (p \, \p_{ z} x-P \, \p_{ z} \xi)(\bar{p} \, \p_{\bar{z}} x-\bar{P} \, \p_{ \bar{z}} \xi) \\
        & + \frac{i \, \sinh^2 (\rho)}{k \, (p\, \bar{p}\, \cosh^2(\rho)+\frac{P\, \bar{P}}{k})} \left( \left( P \, \p_z \xi - p \, \p_z x \right) \, \bar{P} \, \p_{\bar{z}} \theta - \left( \bar{P} \, \p_{\bar{z}} \xi - \bar{p} \, \p_{\bar{z}} x \right) \, P \, \p_z \theta \right) \Bigg].
    \end{split}
\ee
In order to extract a specific target-space geometry from this action, we need to fix a gauge. As long as $p$ and $\bar{p}$ are non-vanishing, a particularly convenient choice is the gauge fixing condition $\xi=0$. By further restricting the model to configurations carrying charges $p=\bar{p}=p'=\bar{p}'$, $P=0$,\footnote{Note that in this particular case the compact boson is charged only in the left-moving sector.} the structure of the action simplifies considerably. In this setting, the corresponding target-space geometry looks
\be\label{eq:3D_wzw_geometry}
    \begin{split}
        ds^2_{(3)} = & \, k \, \tanh^2 (\rho) \, d \theta^2 + k \, d \rho^2 + dx^2 - \frac{i \, \bar{P} \, \tanh^2(\rho)}{p} dx \, d\theta, \\
        B = & - \frac{i \, \bar{P} \, \tanh^2(\rho)}{p} \, dx \wedge d\theta, \\
        \Phi = & \, \Phi_0 - \frac{1}{2} \ln \left( \cosh^2(\rho) \right).
    \end{split}
\ee
The result above is only valid to leading order in the large $k$ limit, where quantum fluctuations are suppressed \cite{Giveon11:1993, Bars:1993, Sfetsos:1993, Tseytlin:1994}. If we keep $p$, $\bar{P}$ and the radius of the $x$-circle fixed as $k\to \infty$, it is justified to dimensionally reduce the $x$-direction and consider the resulting 2D geometry corresponding to a charged black hole 
\be \label{eq:full_2dbh}
    \begin{split}
        ds^2_{(2)} & = k \left[d \rho^2+\tanh^2(\rho) \, \left(1+2 \, a^2  \tanh^2(\rho) \right) \, d\theta^2 \right], \\[5pt]
        A & = - i \, \sqrt{2 \, k}  \, a  \, \tanh^2 (\rho) \, d \theta,  \\[5pt]
        \Phi & = \, \Phi_0 - \frac{1}{2} \ln \left( \cosh^2(\rho) \right),
    \end{split}
\ee
where we defined the convenient parameter
\be \label{eq:extremal_parameter}
    a^2= \frac{\bar{P}^2}{8 \, k \, p^2}.
\ee
This coefficient plays a role analogous to the constant introduced in ($\ref{eq:3dmetric}$), which parametrizes the family of solutions of the three dimensional low-energy string effective action. Note that $a \ll 1$ provided that the dimensional reduction is valid for generic values of $\rho$. However, if one is interested only in the near-horizon physics $\rho \rightarrow 0$, the dimensional reduction remains valid even for generic order one values of $a$. Combining these two observations, we conclude that the dimensional reduction is under control as long as $a^2 \tanh^2(\rho) \ll 1$.

Insofar as this condition holds, we can consistently expand $(\ref{eq:full_2dbh}$) to first order in $a^2$ to obtain 
\begin{equation} \label{eq:metric}
    \begin{split}
        ds^2_{(2)} & = k\left(d \rho^2+\left( \frac{\coth(\rho)}{\coth^2(\rho)-a^2} \right)^{2} d\theta^2\right)+\mathcal{O}\left(a^4 \tanh^4(\rho)\right), \\[5pt]
        A & = - \frac{i \, \sqrt{2 \, k} \, a}{\coth^2 (\rho) - a^2} \,  d \theta + \mathcal{O}\left(a^2 \tanh^2(\rho)\right), \\[5pt]
        \Phi&=\Phi_0-\frac{1}{2}\ln \left( \cosh^2(\rho)-a^2\sinh^2(\rho) \right) + \mathcal{O}\left(a^2 \tanh^2(\rho)\right) ,
    \end{split}
\end{equation}
which precisely agrees with the analytical continuation of the 2D black hole geometry ($\ref{eq:2Dbh}$) derived from the target-space analysis. In the uncharged limit $a=0$, the above expression reduces to Witten's original result \cite{Witten:1991}. Furthermore, dimensionally reducing the two-form $B-$field in ($\ref{eq:3D_wzw_geometry}$) yields a second gauge field analogous to ($\ref{eq:metric}$).

The analysis of the extremal limit $a \to 1$ of the above background closely parallels the discussion in the previous section. It is worth pointing out that, had we constructed the $\frac{SL(2, \mathbb{R}) \times U(1)}{U(1)}$ coset from the outset with charges chosen to satisfy the extremality condition, the resulting black hole would be equivalent to the dimensional reduction of the extremal BTZ black hole \cite{Giveon:2005, Lowe:1994, Strominger:1998}.

\section{Quantum corrections}\label{sec:quantum}

The thermodynamic analysis of the charged 2D black hole outlined in Section \ref{sec:target_analysis} is entirely semiclassical. To go beyond this approximation, one must incorporate stringy effects, which are naturally encoded in the torus partition function of the $\frac{SL(2, \mathbb{R}) \times U(1)}{U(1)}$ WZW model. In the following we initiate the study of this worldsheet partition function, while in Section \ref{sec:free_energy} we will exploit this result to extract the one-loop correction to the quantum black hole entropy. 

\subsection{One-loop worldsheet partition function}\label{sec:partition_fct}

To set the stage, let us recall that the worldsheet description of the charged 2D black hole is provided by the $\frac{SL(2, \mathbb{R}) \times U(1)}{U(1)}$ WZW action ($\ref{eq:action_gauge_wzw}$)
\be
    \begin{split}
        I = \, & \frac{k}{2{\pi}} \int 2 \, d^2\sigma \Bigg[ (\p_z \phi+p \, A_z)(\p_{\bar z} \phi+\bar{p} \, A_{\bar z}) +(\p_z+\p_z \phi+p' \, A_z)\bar v (\p_{\bar z}+\p_{\bar z} \phi+\bar{p}' \, A_{\bar z})v \\
        & \qquad \qquad \qquad +\frac{1}{k} \, (\p_{\bar z} x+ \bar{P} \, A_{\bar z}) (\p_{z} x+P \,A_{z}) \Bigg],
    \end{split}
\ee
where $p$, $\Bar{p}$, $p'$, $\Bar{p}'$, $P$ and $\Bar{P}$ are the coupling constants for the worldsheet gauge field $A$ parameterizing the coset. Since our primary interest lies in the thermal properties of this system, we consider the corresponding grand-canonical ensemble characterized by an inverse temperature $\beta$ and a chemical potential $\mu$. From the target space perspective, the $SL(2,\mathbb{R})$ sector is associated with the three-dimensional hyperboloid $\mathbb{H}_3^+$ with metric
\begin{equation}
    ds^2 = k \left( d\phi^2 +( d v + v d \phi)(d \bar{v} + \bar{v} d \phi) \right).
\end{equation}
Introducing finite temperature and chemical potential amounts to compactifying the Euclidean time direction and turning on a twist along the thermal circle, effectively replacing $H_3^+$ with the corresponding thermal geometry $\mathbb{H}_3^+/ \mathbb{Z}$. Concretely, this is implemented by imposing the following identifications on the coordinates that parametrize $\mathbb{H}_3^+$:
\be \label{eq:identifications}
    \begin{split}
        \left( \phi, v,  \bar{v} \right) \sim \left( \phi + \beta, \, v \, e^{\,i \, \mu \, \beta}, \, \bar{v} \, e^{-i \, \mu \, \beta} \right).
    \end{split}
\ee

The one-loop partition function is computed by formulating the theory on a toroidal worldsheet with modular parameter $\tau=\tau_1 + i \, \tau_2$. The standard coordinate identifications on the torus $z \sim z + 2 \pi\sim z + 2 \, \pi \, \tau$ translate into appropriate worldsheet monodromies of the fields. Consistency with the target space identifications $(\ref{eq:identifications})$ requires \cite{Maldacena:2001}
\be \label{eq:ws_identifications}
    \begin{split}
        \phi(z + 2 \pi) & =\, \phi(z) + \beta \, n, \quad \phi(z + 2 \pi \tau) = \, \phi(z) + \beta \, m, \\
        v(z + 2 \pi) & = \, v(z) \, e^{\, i \, n \,  \mu \, \beta}, \quad v(z + 2 \pi \tau) = \, v(z) \, e^{\, i \, m \, \mu \, \beta},
    \end{split}
\ee
where $n,\,m \in \mathbb{Z}$ label the winding numbers of the string along the two independent cycles of the worldsheet torus. From the target space perspective, these boundary conditions encode both the allowed thermal winding and momentum modes of the string around the Euclidean time circle, as well as the effects of the chemical potential twist. Finally, it is convenient to disentangle the above non-trivial holonomies from the periodic fluctuations by writing
\be \label{eq:holonomy_sl}
    \begin{split}
        \phi = & \,  \hat{\phi} + \beta \, f_{n,m}(z, \bar{z}), \\
        v = &  \, \hat{v} \, e^{\, i \, \mu \, \beta \, f_{n,m}(z,\bar{z})},
    \end{split}
\ee
where $\hat{\phi}$ and $\bar{v}$ are strictly periodic on the torus, while the function
\begin{equation}
    f_{n,m} (z, \bar{z}) = \frac{i}{4 \, \pi \, \tau_2} \left[ z\, (n \, \bar{\tau} - m) - \bar{z} \, (n \, \tau - m) \right]
\end{equation}
captures the winding along the two cycles of the torus and reproduces the worldsheet boundary conditions $(\ref{eq:ws_identifications})$.

On this toroidal worldsheet the action becomes 
\be \label{eq:action_torus}
    \begin{split}
        I =& \frac{k}{2{\pi}} \int 2 \, d^2 \sigma \Bigg[ \left( \p_z \hat{\phi}+pA_z+\frac{U_{n,m}(\bar{\tau}| \beta, 0)}{2\tau_2} \right) \left(\p_{\bar z} \hat{\phi}+\bar{p}A_{\bar z}+\frac{\bar{U}_{n,m}(\tau| \beta,0)}{2\tau_2} \right)\\
        &+ \left(\p_z+\p_z \hat{\phi}+p' A_z+\frac{U_{n,m}(\bar{\tau}| \beta,\mu)}{2\tau_2} \right)\bar{\hat{v}} \left( \p_{\bar z}+\p_{\bar z} \hat{\phi}+\bar{p}'A_{\bar z}+\frac{\bar{U}_{n,m}(\tau|\beta,\mu)}{2\tau_2} \right)\hat{v}\\
        &+\frac{1}{k}(\p_{\bar z} x+\bar{P}A_{\bar z}) (\p_{z} x+PA_{z}) \Bigg],
    \end{split}
\ee
with 
\be
    U_{n,m}(\bar{\tau}|\beta, \mu)=\frac{i\hat{\beta}}{2\pi}(n\bar{\tau}-m), \quad \bar{U}_{n,m}(\tau|\beta, \mu)=-\frac{i\bar{\hat{\beta}}}{2\pi}(n\tau-m), \quad  \hat{\beta} = \beta(1-i\mu).
\ee

To write the partition function in product form we need to Hodge decompose the gauge field as in \cite{Dunne:1998}. For an arbitrary charge configuration, however, the analysis of the associated holonomy becomes rather involved. Therefore, in the remainder of this work we restrict our attention to the simplified case $p=\bar{p}=p'=\bar{p}'=1$, $P=0$ in the canonical ensemble $\mu=0$. In this setting, the Hodge decomposition of the gauge field looks
\begin{equation} \label{eq:hodge_dec}
    \begin{split}
        A= & \partial_z (\eta+i\zeta) \, dz+\partial_{\bar{z}} (\eta-i\zeta) \, d\bar{z} +\frac{1}{2\tau_2} \, \left( U_{n,m}^1 (\bar{\tau}|\alpha,\beta,0) \, dz + \bar{U}^1_{n,m}(\tau|\bar{\alpha},\beta,0) \, d\bar{z} \right),
    \end{split}
\end{equation}
where $\eta$ and $\zeta \sim \zeta + 2 \pi R$ are real bosons, whereas \cite{Hanany:2002, Eguchi:2011}
\begin{equation} \label{eq:gauge}
    \begin{split}
        U_{n,m}^{1}(\bar{\tau}|\alpha,\beta, 0) = & \,U_{n,m}(\bar{\tau}|\beta, 0)  +  \alpha, \quad \bar{U}^{1}_{n,m}(\tau| \bar{\alpha},\beta, 0)= \bar{U}_{n,m}(\tau| \beta, 0)+ \bar{\alpha},\\
        &\alpha = a_1 \bar{\tau}-a_2, \quad a_1, \, a_2 \in (0,1).
    \end{split}
\end{equation}
As will become evident shortly, the path integral over $\eta$ simply gives the volume of the gauge group. We therefore drop any reference to $\eta$ in what follows. While the parameters $a_1$ and $a_2$ can be interpreted as temperature-independent holonomy for the gauge field $\zeta$ \cite{Hanany:2002}, the monodromy of $\zeta$ around the thermal circle looks 
\begin{equation} \label{eq:holonomy_gauge}
    \zeta = \hat{\zeta}- i \beta f_{n,m}(z, \bar{z}),
\end{equation}
where $\hat{\zeta}$ represents the usual periodic quantum fluctuation. Substituting the Hodge decomposition $(\ref{eq:hodge_dec})$ into the action ($\ref{eq:action_torus}$) reveals that $\zeta$ decouples from $\hat{\phi}$ after performing an integration by parts. Moreover, the dependence on $\zeta$ can be absorbed into $\hat{v}$ and $\bar{\hat{v}}$ by the redefinitions 
\be \label{eq:field_red}
    \begin{split}
        \hat{v} & \to e^{\,i \, \zeta } \, \hat{v}, \\
        \bar{\hat{v}} &\to e^{- \, i \, \zeta} \, \bar{\hat{v}}.
    \end{split}
\ee
Keeping track of the change of measure, the torus partition function for the $\frac{SL(2, \mathbb{R}) \times U(1)}{U(1)}$ WZW model is given by
\be \label{eq:wzw_part_funct}
    \begin{split}
        \mathcal{Z}_{\text{wzw}} & =\sum_{n,m \in \mathbb{Z}} \int_{\mathcal{F}} \frac{d^2 \tau}{\tau_2} \frac{|\det(\partial_z \partial_{\bar{z}})|}{4\tau_2^2}\,  \int D\hat{\phi} \, D^2\hat{v} \, Dx \, D \hat{\zeta} \, d^2 \alpha \, \delta(\hat{\zeta}(z_0,\bar{z_0})) \, \mathcal{Z}_{\text{gh}} \, \mathcal{Z}_{\text{int}} \, e^{-I},
    \end{split}
\ee 
where we include the contributions of the usual $bc$ ghost system and the internal CFT
\begin{equation}
    \begin{split}
        \mathcal{Z}_{\text{gh}} & = |\eta(\tau)|^4, \\
        \mathcal{Z}_{\text{int}} & = (q \, \bar{q})^{- \frac{c_{\text{int}}}{24}} \, \sum_{h, \bar{h}} D(h, \bar{h}) \, q^h \, \bar{q}^{\bar{h}}, \qquad q= e^{2 \, \pi \, i \, \tau}.
    \end{split}
\end{equation}
Since the zero mode of $\zeta$ does not enter the Hodge decomposition ($\ref{eq:hodge_dec}$), it has to be omitted both from the corresponding path integral and from the computation of $\det(\p_z \p_{\bar{z}})$. The latter determinant can be equivalently interpreted as the contribution of an additional $bc$ ghost system associated with the gauge fixing of the $U(1)$ gauge symmetry \cite{Hanany:2002}. As usual, integration over the moduli space of the worldsheet torus is restricted to the first fundamental domain, here denoted by $\mathcal{F}$. 

The computation of the above path integral is outlined in Appendix \ref{sec:path_int_ads3}: here we are going to present only the final results. In particular, integration over $\hat{\phi}$, $\hat{\bar{v}}$ and $\hat{v}$ yields the following contribution of the $SL(2, \mathbb{R})$ sector:
\be \label{eq:part_fct_sl}
    \begin{split}
        & \mathcal{Z}_{\mathfrak{sl}_k(2,\mathbb{R})} = \frac{ 4 \pi^2 \beta \sqrt{k-2}}{\sqrt{\tau_2}} \\
        & \frac{e^{-\frac{2 \pi}{\tau_2} \text{Im} (U_{n,m}^{1}(\bar{\tau}| \alpha, \beta, 0)) \text{Im}(\bar{U}^{1}_{n,m}(\tau| \bar{\alpha}, \beta, 0)) - \frac{k \pi}{\tau_2} U^{1}_{n,m}(\bar{\tau}| \alpha, \beta, 0) \bar{U}_{n,m}^{1}(\tau| \bar{\alpha}, \beta, 0) }}{\vartheta_{11}(\bar{\tau}, U_{n,m}^{1}(\bar{\tau}|\alpha, \beta, 0)) \vartheta_{11}(\tau, \bar{U}^{1}_{n,m}(\tau| \bar{\alpha}, \beta, 0))}.
    \end{split}
\ee
This parallels the result presented in \cite{Maldacena:2001}, where the authors computed the thermal partition function of free strings propagating on thermal $AdS_3$. Furthermore, the path integral over $\hat{\zeta}$ can be evaluated by exploiting the results in Appendix \ref{sec:path_int_freefields}, leaving a Gaussian integral over $x$ that can be performed exactly. More explicitly, we get 
\be \label{eq:part_fct_zetax}
    \begin{split}
        \mathcal{Z}_{\hat{\zeta}, x} = & \frac{R}{\tau_2 |\eta(\tau)|^4} \sqrt{k-2 + \frac{ \bar{P}^2}{4}} \sum_{\omega_1, \omega_2 \in \mathbb{Z}} e^{-\frac{\pi}{\tau_2}  \left( | R(\omega_1\bar{\tau} - \omega_2)|^2 + i \bar{P} \bar{U}^1_{n,m}(\tau| \bar{\alpha}, \beta, 0) R (\omega_1\bar{\tau} - \omega_2) \right)}.
    \end{split}
\ee
Overall, the worldsheet partition function reads 
\begin{equation} \label{eq:ws_part_fct}
    \begin{split}
        &\mathcal{Z}_{\text{wzw}} \left( \beta, k, \bar{P}, R \right) = \sum_{n,m \in \mathbb{Z}} \int_{\mathcal{F}} \frac{d^2 \tau}{\tau_2} \, \frac{|\det(\p_z \p_{\bar{z}})|}{4 \tau_2^2} \int d^2 \alpha \, \mathcal{Z}_{\mathfrak{sl}_k(2,\mathbb{R})} \, \mathcal{Z}_{\zeta, x} \, \mathcal{Z}_{\text{gh}} \, \mathcal{Z}_{\text{int}} \\
        = & 2 \pi \beta \sqrt{k-2} \sqrt{k-2 + \frac{ \bar{P}^2}{4}} \sum_{n,m \in \mathbb{Z}} \int_{\mathcal{F}} \frac{d^2 \tau}{\tau_2^{5/2}} \int d^2 \alpha  |\eta(\tau)|^4 (q \bar{q})^{- \frac{c_{\text{int}}}{24}} \sum_{h, \bar{h}} D(h, \bar{h}) q^h \bar{q}^{\bar{h}} \\
        & (2 \pi R) \sum_{\omega_1, \omega_2 \in \mathbb{Z}} e^{-\frac{\pi}{\tau_2}  \left( | R(\omega_1\bar{\tau} - \omega_2)|^2 + i \bar{P} \bar{U}^1_{n,m}(\tau|\bar{\alpha}, \beta, 0) R (\omega_1\bar{\tau} - \omega_2) \right)} \\
        & \frac{e^{-\frac{2 \pi}{\tau_2} \text{Im} \left( U_{n,m}^{1}(\bar{\tau}| \alpha, \beta,0) \right) \text{Im} \left( \bar{U}^{1}_{n,m}(\tau| \bar{\alpha}, \beta, 0) \right) - \frac{k \pi}{\tau_2} U^{1}_{n,m}(\bar{\tau}| \alpha, \beta, 0) \bar{U}_{n,m}^{1}(\tau| \bar{\alpha}, \beta, 0)}}{\vartheta_{11}(\bar{\tau}, U_{n,m}^{1}(\bar{\tau}|\alpha, \beta, 0)) \vartheta_{11}(\tau, \bar{U}^{1}_{n,m}(\tau| \bar{\alpha}, \beta, 0))}.
    \end{split}
\end{equation}
The above expression is modular invariant, as explicitly verified in Appendix \ref{sec:modular_inv}, providing a nontrivial consistency check of the construction. Moreover, this result is consistent with the partition function of the $\frac{SL(2,\mathbb{R})}{U(1)}$ uncharged black hole derived in \cite{Hanany:2002} when $\beta=2 \pi i$, $\bar{P}=0$ and the contribution from the compact $U(1)$ boson is neglected. The apparent discrepancy arises from our use of different conventions for the holonomies $(\ref{eq:holonomy_sl})$ and $(\ref{eq:holonomy_gauge})$, as well as a different expression for the Ray–Singer torsion \cite{Ray:1973} used in the computation of ($\ref{eq:part_fct_sl})$, which in our case is not periodic in the holonomy of the fields.

As outlined in Section \ref{sec:ws_analysis}, the charged 2D black hole arises from the dimensional reduction of the $\frac{SL(2, \mathbb{R}) \times U(1)}{U(1)}$ WZW model. Accordingly, its worldsheet partition function is obtained by isolating the zero-mode contribution associated with the compact $U(1)$-direction in ($\ref{eq:ws_part_fct}$). Since the worldsheet partition function encodes the one-loop contribution to the target-space free energy \cite{Maldacena:2001}, the corresponding target-space partition function takes a form analogous to ($\ref{eq:formula_target_part}$), namely
\begin{equation}
    Z_{2D} \, = \, \exp \left( S_0 \,+ \,\mathcal{Z}_{2D}\, + \, ... \right).
\end{equation}
Here $Z_{2D}$ is the 2D target-space partition function, $\mathcal{Z}_{2D}$ is the 2D worldsheet partition function, and $S_0$ is the classical entropy ($\ref{eq:classical_entropy}$). From the string theory perspective, the latter contribution should be encoded in the genus-zero worldsheet partition function. To the best of our knowledge, no such computation has been performed in the context of bosonic string theory. In the next section, we use $\mathcal{Z}_{\text{wzw}}$ to extract the temperature dependence of the one-loop corrections to the black hole quantum entropy.

\subsection{One-loop target-space free energy}\label{sec:free_energy}

As anticipated at the end of the previous section, the worldsheet partition function provides direct access to the one-loop contribution to the target-space free energy \cite{Maldacena:2001}. More specifically, the worldsheet partition function can be written as  
\be \label{eq:free_en_target_sp}
    \begin{split}
         \mathcal{Z}_{\text{wzw}} = - \beta \, \sum_{m=1}^{\infty} \, f_{\text{wzw}} (m \, \beta),
    \end{split}
\ee
where $f_{\text{wzw}} (m \, \beta)$ encodes the contribution from the $m$-string Fock space, and the sum over $m$ accounts for all possible multi-string states. It is therefore sufficient to compute the single-string contribution and systematically incorporate multi-string sectors through ($\ref{eq:free_en_target_sp}$).

In the original form of the partition function $(\ref{eq:ws_part_fct})$, the sum runs over both winding numbers $m, n \in \mathbb{Z}$ and the integration over the torus moduli space is restricted to the first fundamental domain $\mathcal{F}$. It is well known, however, that the sum over winding sectors $(m,n)\neq (0,0)$ can be unfolded into a sum over $m \geq 0$ fixing $n=0$, provided the integration region for the modular parameter is enlarged to the strip $\tau_2>0$, $|\tau_1|<1/2$. Since the $m=0$ term represents a divergent contribution to the zero temperature vacuum energy, we will discard it in the following. The single-string contribution then looks
\begin{equation}\label{eq:fm}
    \begin{split}
        f_{\text{wzw}} (\beta)=  & \, C \int d^2 \alpha \int_{-\frac{1}{2}}^{\frac{1}{2}} d\tau_1 \int_0^\infty \frac{d \tau_2}{\tau_2^{5/2}} e^{4 \pi \tau_2 \left( 1-\frac{1}{4(k-2)} \right)}\sum_{h, \bar{h}} D(h,\bar{h}) q^h \bar{q}^{\bar{h}} \\
         & \sum_{\omega_1, \omega_2 \in \mathbb{Z}} R \,  e^{-\frac{\pi}{\tau_2}  \left( | R(\omega_1\bar{\tau} - \omega_2)|^2 + i \bar{P} \bar{U}^{1}_{0,1}(\tau| \bar{\alpha}, \beta, 0)) R (\omega_1\bar{\tau} - \omega_2) \right)} \\
        & \frac{e^{\frac{2 \pi}{\tau_2} \text{Im}\left(\alpha-\frac{i \beta}{2\pi}\right)^2  - \frac{k\pi}{\tau_2} |\alpha -i\frac{\beta}{2 \pi}|^2}}{\Big|\sinh \left( \frac{\beta+2 \pi i \alpha }{2} \right)\Big|^2} \Bigg| \prod_{r=1}^\infty \frac{1- e^{2 \pi i \tau r}}{\left( 1- e^{  -\beta - 2 \pi i \alpha - 2 \pi i \bar{\tau} r} \right)  \left( 1- e^{\beta + 2 \pi i \alpha - 2 \pi i \bar{\tau} r} \right)} \Bigg|^2,
    \end{split}
\end{equation}
where the normalization constant $C$ is
\be
    C=  \, \pi^2 \, \sqrt{k-2} \, \sqrt{k-2+\frac{\bar{P}^{2}}{4}}.
\ee

To make contact with the dimensionally reduced theory, we perform a Poisson resummation over the winding modes associated with the compact $U(1)$ direction:
\be \label{eq:poisson_resum}
    \begin{split}
        & \sum_{\omega_1, \omega_2 \in \mathbb{Z}}  R \, e^{-\frac{\pi}{\tau_2}  \left( | R(\omega_1\bar{\tau} - \omega_2)|^2 + i \bar{P} \bar{U}^{1}_{0,1}(\tau| \bar{\alpha}, \beta, 0) R (\omega_1\bar{\tau} - \omega_2) \right)}\\
        =&\sum_{\omega, p \in \mathbb{Z}} \tau_2^{\frac{1}{2}} \, e^{- \pi \tau _2 \left(\frac{p^2}{R^2}+ \omega^2 R^2\right)+2 \pi i \tau _1 \omega p +\pi \bar{P} \bar{U}^{1}_{0,1}(\tau| \bar{\alpha}, \beta, 0)   \left(\frac{p}{R} + \omega R +\frac{\bar{P} \bar{U}^{1}_{0,1}(\tau| \bar{\alpha}, \beta, 0)}{4 \tau _2}\right)}.
    \end{split}
\ee
To get to this formula we have performed Poisson resummation on $\omega_2$ and renamed $\omega_1$ to $\omega$. The single-string free energy associated with the 2D black hole is then obtained by setting $\omega=p=0$.

In order to analytically compute ($\ref{eq:fm}$) we follow the discussion outlined in \cite{Maldacena:2001} (see \cite{Ferko:2025} for a more formal treatment). To start, we impose the identification $(\ref{eq:gauge})$ $\alpha = a_1\bar{\tau}-a_2$ to find that the poles of ($\ref{eq:fm}$) in the complex $\tau_2-$plane lie at $\tau_2=\frac{\beta}{2 \pi (w+a_1)}$, where $w \in [0, \infty)$ plays the usual role of spectral flow parameter associated with the $\mathfrak{sl}_k(2,\mathbb{R})$ representations. Accordingly, we divide the domain of $\tau_2$ integration into cells
\begin{equation} \label{eq:cells}
    [ 0, \infty) = \bigcup_{w=0}^\infty \frac{\beta}{2 \pi} \left[ \frac{1}{w+a_1+1}, \frac{1}{w+a_1} \right].
\end{equation}
Within each cell, the first factor in the denominator of the infinite product in ($\ref{eq:fm}$) can be expanded as
\begin{equation}
    \frac{1}{1-e^{\beta -2 \pi \bar{\alpha} + 2 \pi i \tau}} =
    \begin{dcases}
        \sum_{\ell=0}^\infty e^{\, \ell(\beta -2 \pi \bar{\alpha} + 2 \pi i w \tau)} \qquad \qquad \quad \tau_2 > \frac{\beta}{2 \pi (w+a_1)}, \\
        -\sum_{\ell=0}^\infty e^{-(\ell+1)(\beta -2 \pi \bar{\alpha} + 2 \pi i w \tau)} \qquad \tau_2 < \frac{\beta}{2 \pi (w+a_1)}.
    \end{dcases} 
\end{equation}
Analogous expansions hold for the second factor and the antiholomorphic contributions. Combining these results, the single-string free energy for the dimensionally reduced theory looks 
\be \label{eq:free_2d} 
    \begin{split}
        f_{\text{2D}} (\beta) &= \, 4 C \sum_{h, \bar{h}} D(h,\bar{h}) \sum_{w=0}^\infty \sum_{\ell, \bar{\ell} =0}^\infty \sum_{N, \bar{N} \in \mathbb{Z}}  \int d^2 \alpha \int_{-\frac{1}{2}}^{\frac{1}{2}} d \tau_1 \int_{\frac{\beta}{2 \pi(w+a_1+1)}}^{\frac{\beta}{2 \pi (w+a_1)}} \frac{d \tau_2}{\tau_2^2} e^{4 \pi \tau_2 \left( 1-\frac{1}{4(k-2)} \right)} \\
        & e^{- \frac{\pi k}{\tau_2} |\alpha|^2 + \frac{ 2 \pi \alpha_2^2}{\tau_2} - \left( k-2 -\frac{\bar{P}^2}{4} \right)\frac{\beta^2}{4 \pi \tau_2} + \left( \left(k-2 -\frac{\bar{P}^2}{4} \right)\alpha_2 - \frac{i \alpha_1 \bar{P}^2}{4} \right) \frac{\beta}{\tau_2} - \frac{\pi \bar{\alpha}^2 \bar{P}^2}{4 \tau_2} } \\
       & e^{-\left( \beta -2 \pi i \bar{\alpha}\right) \left( \frac{1}{2}+w+\ell \right) - \left( \beta + 2 \pi i \alpha \right) \left( \frac{1}{2}+w+\bar{\ell} \right)} e^{2 \pi i \tau_1 \left( N +h -\bar{N} - \bar{h} \right)} e^{-2 \pi \tau_2 \left( N + h + \bar{N} + \bar{h} -w(w+1) \right)},
    \end{split}
\ee
where the parameters $N$, $\bar{N}$ are associated to the excitation level of the $\mathfrak{sl}_k(2,\mathbb{R})$ representations. Here we refrain from explicitly performing the substitution $\alpha = a_1\bar{\tau}-a_2$ in order to keep the above expression more compact. 

Since we are interested in the large-$\beta$ limit of the above expression, $(\ref{eq:cells})$ shows that the dominant integration region for $\tau_2$ is roughly $\Big[\frac{\beta}{2\pi}, \infty \Big)$, corresponding to the sector of the $\mathfrak{sl}_k(2, \mathbb{R})$ algebra with no spectral flow $(w=0)$. This sector is dominated by the short-string states associated with discrete representations of $\mathfrak{sl}_k(2, \mathbb{R})$ \cite{Maldacena1:2001}. It is therefore natural to restrict the computation of $f_{\text{2D}}(\beta)$ to the contribution from short strings. 

Despite this restriction, we are unable to find a closed-form result, the main technical problem being the integration over the gauge-field holonomy $a_1$. The best we can do is to develop a recursive procedure that determines corrections to the free energy order by order in $1/\beta$. Since we are ultimately interested in the low-temperature regime of these corrections, we will retain and discuss only the leading-order term. As explicitly detailed in Appendix \ref{sec:euclidean}, in this limit the worldsheet partition function for the 2D black hole in the canonical ensemble looks
\begin{equation} \label{eq:2d_part_funct}
    \begin{split}
        \mathcal{Z}_{2D}\Big|_{\text{short strings}} = & \, 2 \pi^2 \, \sqrt{k-2} \, \sqrt{k-2+\frac{\bar{P}^{2}}{4}} \int_0^1 da_1 \, \sum{\vphantom{\sum}}' D'(h,\bar{h}, N, \bar{N}, \ell, \bar{\ell}, w; a_1) \\
        & \sqrt{\beta} \, \text{Li}_{1/2}\left( e^{-\beta \left( 1+\ell+\bar{\ell}+2w + 2 \, \sqrt{\left(k-2-\frac{\bar{P}^2}{4}\right) \left( N + h - \frac{w(w+1)}{2} -1 + \frac{1}{4(k-2)} \right)} \, \right) } \right),
    \end{split}
\end{equation}
where $\text{Li}_{s}(z)$ is the polylogarithmic function of order $s$ \cite{Polylog} 
\begin{equation}
    \text{Li}_{s}(z) = \sum_{n=1}^\infty \frac{z^n}{n^s},
\end{equation}
and 
\begin{equation}\label{eq:coeff_d}
     D'(h,\bar{h}, N, \bar{N}, \ell, \bar{\ell}, w; a_1) = \sqrt{\frac{2 \pi( w + a_1+1)}{N + h -\frac{w(w+1)}{2} -1 + \frac{1}{4(k-2)}}} \, D(h,\bar{h}).
\end{equation}
Moreover, the primed sum runs over $h$, $\bar{h}$, $N$, $\bar{N}$, $\ell$, $\bar{\ell}$, $w$ and it is constrained by the level matching condition $N+h=\bar{N}+\bar{h}$ and the constraint derived in (\ref{eq:second_constr}):
\begin{equation}\label{eq:bound_states}
    \frac{k-2-\frac{\bar{P}^{2}}{4}}{4} (w + a_1)^2 < N + h -\frac{w(w+1)}{2} -1 + \frac{1}{4(k-2)} <  \frac{k-2-\frac{\bar{P}^{2}}{4}}{4} (w + a_1+1)^2.
\end{equation}
Although this result is not in closed form, it allows one to extract the temperature dependence of the leading one-loop correction to the quantum entropy of the black hole. Before proceeding, it is useful to comment on some general features of the result.

The structure of the exponential in (\ref{eq:2d_part_funct}) closely resembles the expression found in \cite{Maldacena:2001}; specifically, the energy spectrum of short strings on $AdS_3$ is recovered in the limit $\bar{P} \to 0$. The main difference lies in the bound ($\ref{eq:bound_states}$), which here depends on the holonomy parameter $a_1$. While this parameter is integrated over in the full partition function, explicitly evaluating this integral is unnecessary for our current focus on temperature dependence; it becomes relevant only when inferring the full 2D black hole spectrum. A detailed analysis of the black hole spectrum as well as of the long-string contribution will be presented in a future work. 

As explicitly shown in Appendix \ref{sec:euclidean}, the validity of the above result hinges on the condition $\left( k-2-\frac{\bar{P}^2}{4} \right) > 0$. Expressed in terms of the black hole temperature $T = \frac{1-a^2}{2 \pi \sqrt{k}}$ from (\ref{eq:classical_temp}), this bound translates to
\begin{equation}\label{eq:temp_limit}
    T \gtrsim \frac{1}{4 \pi \sqrt{k}},
\end{equation}
where we have used the relation $a^2 = \frac{\bar{P}^2}{8k}$ from (\ref{eq:extremal_parameter}) and retained the dominant term in a $1/k$ expansion. Consequently, our result holds strictly away from extremality (that corresponds to the limit $\bar{P}^2 \to 8k$). As noted below \eqref{eq:long_string}, the threshold at which $\left( k-2-\frac{\bar{P}^2}{4} \right)$ turns negative corresponds to the point where the long-string contribution to the free energy diverges. A definitive confirmation of this claim, however, requires a rigorous analysis of the long-string spectrum, which lies beyond the scope of the current approach.

As discussed at the end of Section \ref{sec:partition_fct}, the worldsheet partition function ($\ref{eq:2d_part_funct}$) corresponds to the one-loop correction to the quantum entropy of the 2D black hole. Since the polylogarithmic function behaves as $\text{Li}_{1/2}(z) \approx z$ for small values of its argument $z \ll 1$, the near-extremal quantum entropy in the canonical ensemble takes the form 
\begin{equation}\label{eq:part_func_extremal}
    \begin{split}
        & \ln(Z_{2D}) =  \pi  \sqrt{k} \, Q + \frac{\pi^2 k \, Q}{\beta} +2 \pi^2 \, \sqrt{k-2} \, \sqrt{k-2+\frac{\bar{P}^{2}}{4}} \int_0^1 da_1 \\
        & \sum{\vphantom{\sum}}' D'(h,\bar{h}, N, \bar{N}, \ell, \bar{\ell}, w; a_1) \sqrt{\beta} \,  e^{-\beta \left( 1+\ell+\bar{\ell}+2w + 2\sqrt{\left(k-2-\frac{\bar{P}^2}{4}\right) \left( N + h -\frac{w(w+1)}{2} -1 + \frac{1}{4(k-2)} \right)} \, \right)},
    \end{split}
\end{equation}
where we have also included the classical contribution derived in ($\ref{eq:near_ext_entropy}$). The parameter $Q$ is the overall black hole charge given in ($\ref{eq:mass}$), which is fixed in the canonical ensemble.

Two important physical remarks are in order. First, the temperature-dependent logarithmic correction obtained via the target-space analysis in Section \ref{sec:target_analysis} is absent from \eqref{eq:part_func_extremal}. That correction arose from a dimensional reduction to JT gravity, which inherently assumes an infinitely long near-horizon throat. Its disappearance here signals that no such throat develops, indicating that the solution behaves as a small black hole. 

Secondly, by recalling that $\beta$ is related to the parameters $(k, a)$ via \eqref{eq:classical_temp}
\begin{equation}
    \beta = \frac{2 \pi \sqrt{k}}{1-a^2},
\end{equation}
the near-extremal expansion in \eqref{eq:part_func_extremal} remains well-defined provided 
\begin{equation}
    \beta \gg 0, \qquad \beta \lesssim 4 \pi \sqrt{k}.
\end{equation}
Here, the second condition, analogous to \eqref{eq:temp_limit}, ensures that $ \left( k - 2 - \frac{\bar{P}^2}{4} \right) > 0$. The self-consistency of the expansion in \eqref{eq:part_func_extremal} is then controlled by $k$: for sufficiently large values of $k$, both conditions are satisfied simultaneously and the result is fully trustworthy.

This threshold behavior admits a natural physical interpretation in terms of the black hole/string transition. For uncharged 2D black holes, it has been shown \cite{Giveon_transition:2005, Kutasov:2005, Giveon:2007} that as the black hole temperature increases, the geometry develops large thermal fluctuations across an expanding region around the horizon. When the temperature approaches the Hagedorn threshold \cite{Atick:1988}, the black hole becomes indistinguishable from a gas of fundamental strings. In the dual Euclidean geometry, this transition is mediated by a non-zero condensate of the closed string tachyon wound around the Euclidean time direction, whose mass winds down toward zero as the system heats up.

The analogous transition for charged 2D black holes has been explored in \cite{Giveon:2006}. Crucially, the string/black hole transition for charged black holes can occur at temperatures arbitrarily low compared to the Hagedorn temperature. This latter observation offers an explanation for why our target-space partition function ($\ref{eq:part_func_extremal}$), which accounts only for the short-string contribution, contains a signature of the black hole/string transition, which is inherently driven by the long-string density of states. Ultimately, this interplay between the Hagedorn transition and the low-temperature limit of a charged black hole may mirror the mechanism discussed in \cite{Chen:2021}, where the extremal limit of the charged Horowitz-Polchinski string \cite{Horowitz:1998} is shown to map precisely to the Hagedorn transition of the corresponding neutral solution.

\section{Conclusion}\label{sec:conclusion}

The goal of this paper was to compute the one-loop correction to the near-extremal entropy of the charged 2D black hole introduced in \cite{McGuigan:1992}. At the semiclassical level, this black hole can be interpreted either as a solution of the two-dimensional Einstein–Maxwell theory or as arising from the dimensional reduction of a class of solutions of the three-dimensional low-energy string effective action. In principle, one could therefore incorporate the quantum fluctuations around this background by evaluating the corresponding gravitational path integral and extracting from it the one-loop correction to the near-extremal entropy. While we did not perform this explicit calculation, we derived the standard temperature-dependent logarithmic correction to the entropy of our black hole by adapting the analysis of the four-dimensional Reissner-Nordstrom black hole in asymptotically Anti-de Sitter spacetime detailed in \cite{Iliesiu:2021}.

However, the existence of a worldsheet description of this 2D black hole in terms of the $\frac{SL(2,\mathbb{R}) \times U(1)}{U(1)}$ WZW model allows for a genuinely string-theoretic computation. In this work we therefore approached the problem by evaluating the torus partition function, which encodes the one-loop contribution in the worldsheet expansion. Owing to the technical complexity of the calculation, we were unable to obtain a closed-form expression for the full result. Instead, we restricted our analysis to the contribution arising from short-string states, which are expected to dominate in the large-$\beta$ limit. For similar technical reasons, we were also unable to determine the complete spectrum of the black hole. This problem is left for future work, where we also plan to provide a more systematic treatment of the long-string sector.

Within this approximation we were nevertheless able to determine the temperature dependence of the worldsheet partition function and extract from it the leading one-loop correction to the near-extremal quantum entropy of the black hole. The resulting expression $(\ref{eq:part_func_extremal})$ is valid provided $\beta \gg 0$ and $\beta \lesssim 4 \pi \sqrt{k}$. We speculate that this threshold behavior of the partition function can be interpreted through the lens of the black hole/string transition.

\section*{Acknowledgments}\addcontentsline{toc}{section}{Acknowledgments}

I would like to thank my Ph.D advisor, Prof. Mukund Rangamani, for guiding me through the preparation of this paper and for providing many valuable suggestions. I am also grateful to Ziyi Li for pointing out a crucial mistake in an earlier version of this paper regarding the semiclassical analysis in Section \ref{sec:target_analysis}.

\newpage

\appendix

\section{Partition function of free field theories}\label{sec:path_int_freefields}

In this appendix we collect the expressions for the torus partition functions of free field theories that will be needed in the subsequent discussion. To set the stage, we remind the reader of the basic conventions for the definition of a torus, parameterized by complex coordinates $(z,\bar{z})$ and characterized by the modular parameter $\tau=\tau_1 + i \tau_2$ \cite{Polchinski:2007}:  
\be
    z\sim z+2 \, \pi \sim z+2 \, \pi \, \tau, \quad \int d^2 \sigma=4 \, \pi^2 \, \tau_2.
\ee
For a real non-compact scalar field $\zeta$, the partition function looks
\be \label{eq:det}
    Z = \int D\zeta \ \delta(\zeta(z_0,\bar{z_0})) e^{-\frac{1 }{2{\pi}}
    \int 2d^2\sigma \  \p_z \zeta \p_{\bar{z}} \zeta}=\bigg( \frac{4 \tau_2}{\det(-\p_z \p_{\bar{z}})}\bigg)^{\frac{1}{2}}=\frac{1}{( \tau_2)^{\frac{1}{2}}|\eta(\tau)|^2}.
\ee
On the other hand, the partition function of a compact real scalar field $\zeta \sim \zeta+2\pi R$ coupled to a background gauge field $A$ is given by \cite{Chen:2013}\footnote{The map from here to \cite{Chen:2013} is given by $z \to 2 \pi z, R=\sqrt{2}/g$.}
\be\label{CompactBoson}
    \begin{split}
        Z[A_z,A_{\bar{z}}] & = \int D\zeta \, e^{-\frac{1 }{2{\pi}}
        \int 2 \, d^2\sigma \  (\p_z \zeta+A_z)( \p_{\bar{z}} \zeta+A_{\bar{z}})} \\
        & = Z^{z}[A_z,A_{\bar{z}}] \, Z^{nz}[A_z,A_{\bar{z}}] \, e^{-\frac{1 }{2{\pi}}
        \int 2 \, d^2 \sigma \  A_z \, A_{\bar{z}}}.
    \end{split}
\ee
If we decompose the gauge field in the Fourier basis
\be\label{Fourier}
    A_z=\sum_{m,n} A_z^{m,n} \, e^{\frac{1}{2\tau_2} \, ( \, m \, ( \, \bar{z} \, \tau-z \, \bar{\tau} \, )+n \, (\, z-\bar{z} \, ))},
\ee
we can write the explicit expressions for $Z^{z}$ and $Z^{nz}$ as
\begin{equation}
    \begin{split}
        Z^{nz} [A_z,A_{\bar{z}}] & = \frac{e^{-\pi\tau_2\sum'_{m,n}(A_z-A_{\bar{z}})^{m,n}(A_z-A_{\bar{z}})^{-m,-n}}}{( \tau_2)^{\frac{1}{2}}|\eta(\tau)|^2} e^{\sum'_{m,n}\frac{m\pi \tau_2 A_{z}^{m,n}A_{z}^{-m,-n}}{\Pi_{z}^{m,n}}} e^{\sum'_{m,n}\frac{m\pi \tau_2 A_{\bar{z}}^{m,n}A_{\bar{z}}^{-m,-n}}{\Pi_{\bar{z}}^{m,n}}}, \\
        Z^{z} [A_z,A_{\bar{z}}] & = ( \tau_2)^{\frac{1}{2}} \sum_{n,w}q^{\frac{1}{4}(\frac{n}{R}+w R)^2} \bar{q}^{\frac{1}{4}(\frac{n}{R}-w R)^2}e^{-\pi \tau_2(A_z^{0,0}-A_{\bar{z}}^{0,0})^2-2\pi\tau_2 (\frac{n}{R}+wR)A^{0,0}_{z}-2\pi \tau_2 (\frac{n}{R}-wR)A^{0,0}_{\bar{z}}},
    \end{split}
\end{equation}
with
\begin{equation}
     \Pi_z^{m,n}=i \, \frac{( \, m \, \bar{\tau}-n \, )}{2 \, \tau_2}.
\end{equation}
Here $m, \,n, \, w \in \mathbb{Z}$ and a prime symbol indicates that the sum does not include the zero mode $(m,n)=(0,0)$.

\section{$\frac{SL(2, \mathbb{R}) \times U(1)}{U(1)}$ coset partition function}\label{sec:path_int_ads3}

In this appendix, we provide the technical details underlying the computation of the worldsheet partition function of the $\frac{SL(2, \mathbb{R}) \times U(1)}{U(1)}$ WZW model discussed in Section \ref{sec:partition_fct}. 

Our starting point is ($\ref{eq:wzw_part_funct}$), which we reproduce here for convenience 
\be 
    \begin{split}
        \mathcal{Z}_{\text{wzw}} & =\sum_{n,m \in \mathbb{Z}} \int_{\mathcal{F}} \frac{d^2 \tau}{\tau_2} \frac{|\det(\partial_z \partial_{\bar{z}})|}{4\tau_2^2}\,  \int D\hat{\phi} \, D^2\hat{v} \, Dx \, D \hat{\zeta} \, d^2 \alpha \, \delta(\zeta(z_0,\bar{z_0})) \, \mathcal{Z}_{\text{gh}} \, \mathcal{Z}_{\text{int}} \, e^{-I},
    \end{split}
\ee 
where the corresponding action looks (\ref{eq:action_torus})
\be \label{eq:action_torus_2}
    \begin{split}
        I =& \frac{k}{2{\pi}} \int 2 \, d^2 \sigma \Bigg[ \left( \p_z \hat{\phi}+pA_z+\frac{U_{n,m}(\bar{\tau}| \beta, 0)}{2\tau_2} \right) \left( \p_{\bar z} \hat{\phi}+\bar{p}A_{\bar z}+\frac{\bar{U}_{n,m}(\tau| \beta,0)}{2\tau_2} \right)\\
        &+ \left(\p_z+\p_z \hat{\phi}+p' A_z+\frac{U_{n,m}(\bar{\tau}| \beta,\mu)}{2\tau_2} \right)\bar{\hat{v}} \left( \p_{\bar z}+\p_{\bar z} \hat{\phi}+\bar{p}'A_{\bar z}+\frac{\bar{U}_{n,m}(\tau|\beta,\mu)}{2\tau_2} \right)\hat{v}\\
        &+\frac{1}{k}(\p_{\bar z} x+\bar{P}A_{\bar z}) (\p_{z} x+PA_{z}) \Bigg].
    \end{split}
\ee
Although the main text focuses on the simplified case
\begin{equation} \label{eq:scenario}
    p=p'=\bar{p}=\bar{p}'=1, \qquad P=0, \qquad \mu=0,
\end{equation}
in this section we carry out the analysis in full generality when possible.

We begin by mentioning that the term $\frac{|\det(\partial_z \partial_{\bar{z}})|}{4\tau_2^2}$ can be computed by using (\ref{eq:det}). Then, we perform the integral over the fields parameterizing the $SL(2, \mathbb{R})$ sector. The core of the computation uses the Ray-Singer analytic torsion arising from the path integral over $\hat{v}$ and $\hat{\bar{v}}$. After imposing the Hodge decomposition $(\ref{eq:hodge_dec})$ and (\ref{eq:gauge}) of the gauge field and the field redefinitions (\ref{eq:field_red}), the relevant piece of the action is
\begin{equation} \label{eq:gauged_v_vbar}
    I_{\hat{v},\hat{\bar{v}}} = \frac{k}{2{\pi}} \int 2 \, d^2 \sigma \left(\p_z+\p_z \hat{\phi}+\frac{U_{n,m}^{p'}(\bar{\tau}| \alpha, \beta,\mu)}{2\tau_2} \right)\hat{\bar{v}} \left( \p_{\bar z}+\p_{\bar z} \hat{\phi}+\frac{\bar{U}^{\bar{p}'}_{n,m}(\tau|\bar{\alpha}, \beta,\mu)}{2\tau_2} \right)\hat{v},
\end{equation}
where
\be
    U_{n,m}^{p'}(\bar{\tau}|\alpha,\beta, \mu) = U_{n,m}(\bar{\tau}|\beta, \mu)+p'\alpha, \quad \bar{U}^{\bar{p}'}_{n,m}(\tau| \bar{\alpha},\beta, \mu)= \Bar{U}_{n,m}(\tau| \beta, \mu)+\Bar{p}' \Bar{\alpha},
\ee
represent the generalization of (\ref{eq:gauge}). Note that the action is quadratic in $\hat{v}$ and $\hat{\bar{v}}$. Since we can disentangle the $\hat{\phi}$-dependence by a chiral rotation, the path integral over $\hat{v}$ and $\hat{\bar{v}}$ becomes the regularized determinant of the Laplacian on the space of functions that have nontrivial holonomies around the cycles of the worldsheet torus. By using the $\zeta-$function regularization, this regularized determinant is known as the Ray-Singer analytic torsion \cite{Ray:1973}: 
\be \label{eq:torsion}
    \begin{split}
        &  \det \left[- \left( \p + \p \hat{\phi}  + \frac{U_{n,m}^{p'}(\bar{\tau}| \alpha,\beta, \mu)}{2 \tau_2} \right)^{-1} \left( \bar{\p} + \bar{\p} \hat{\phi} + \frac{\bar{U}^{\bar{p}'}_{n,m}( \tau| \bar{\alpha},\beta, \mu)}{2 \tau_2} \right)^{-1} \right] \\
        & =  e^{\frac{2}{\pi} \int d^2 \sigma \p \hat{\phi} \bar{\p} \hat{\phi}} \det \left[- \left( \p + \frac{U_{n,m}^{p'}(\bar{\tau}| \alpha,\beta, \mu)}{2 \tau_2} \right)^{-1} \left( \bar{\p} + \frac{\bar{U}^{\bar{p}'}_{n,m}( \tau| \bar{\alpha},\beta, \mu)}{2 \tau_2} \right)^{-1} \right] \\
        & = e^{\frac{2}{\pi} \int d^2 \sigma \p \hat{\phi}  \bar{\p} \hat{\phi}} \, \frac{4 \pi^2 | \eta(\tau)|^2 e^{-\frac{2 \pi}{\tau_2} \text{Im} (U_{n,m}^{p'}(\bar{\tau}| \alpha, \beta, \mu)) \text{Im}(\bar{U}^{\bar{p}'}_{n,m}(\tau| \bar{\alpha}, \beta, \mu))}}{\vartheta_{11}(\bar{\tau}, U_{n,m}^{p'}(\bar{\tau}| \alpha, \beta, \mu)) \vartheta_{11}(\tau, \bar{U}^{\bar{p}'}_{n,m}(\tau| \bar{\alpha}, \beta, \mu))}.
    \end{split}
\ee
On the other hand, the effective action that enters the path integral over $\hat{\phi}$ is obtained by assembling the relevant terms in (\ref{eq:action_torus_2}), (\ref{eq:gauged_v_vbar}) and (\ref{eq:torsion})
\be
    \begin{split}
        I_{\hat{\phi}} = & \frac{k}{2 \pi} \int 2 \, d^2 \sigma \left( \p \hat{\phi} + \frac{U_{n,m}(\bar{\tau}| \beta, 0) + p \alpha}{2 \tau_2} \right) \left( \bar{\p} \hat{\phi} + \frac{\bar{U}_{n,m} (\tau| \beta, 0)+ \bar{p} \bar{\alpha}}{2 \tau_2} \right) \\
        & - \frac{2}{\pi} \int d^2 \sigma \p \hat{\phi} \bar{\p} \hat{\phi} \\
        = & \frac{k \pi}{\tau_2} \left( \frac{\beta}{2 \pi}(n \tau -m) + i \bar{p} \bar{\alpha} \right) \left( \frac{\beta}{2 \pi}(n \bar{\tau} -m) - i p \alpha \right) + \frac{k-2}{\pi} \int d^2 \sigma \p \hat{\phi} \bar{\p} \hat{\phi},
    \end{split}
\ee
where the second equality follows from the periodicity of $\hat{\phi}$. The corresponding path integral yields 
\be
    \int D \hat{\phi} \, e^{- I_{\hat{\phi}}} = \frac{\beta \, \sqrt{k-2}}{\sqrt{\tau_2} \, |\eta(\tau)|^2} e^{ - \frac{k \pi}{\tau_2} \left( \frac{\beta}{2 \pi}(n \tau -m) + i \bar{p} \bar{\alpha} \right) \left( \frac{\beta}{2 \pi}(n \bar{\tau} -m) - i p \alpha \right)}.
\ee
Hence, the contribution to the worldsheet partition function coming from the $SL(2, \mathbb{R})$ sector of the WZW model is \cite{Maldacena:2001}
\be
    \begin{split}
        & \mathcal{Z}_{\mathfrak{sl}_k(2, \mathbb{R})} = 4 \, \pi^2 \, \frac{\beta \, \sqrt{k-2}}{\sqrt{\tau_2}} \\
        & \frac{e^{-\frac{2 \pi}{\tau_2} \text{Im} (U_{n,m}^{p'}(\bar{\tau}| \alpha, \beta, \mu)) \text{Im}(\bar{U}^{\bar{p}'}_{n,m}(\tau| \bar{\alpha}, \beta, \mu)) - \frac{k \pi}{\tau_2} U^{p}_{n,m}(\bar{\tau}| \alpha, \beta, 0) \bar{U}_{n,m}^{\bar{p}}(\tau| \bar{\alpha}, \beta, 0) }}{\vartheta_{11}(\bar{\tau}, U_{n,m}^{p'}(\bar{\tau}|\alpha, \beta, \mu)) \vartheta_{11}(\tau, \bar{U}^{\bar{p}'}_{n,m}(\tau| \bar{\alpha}, \beta, \mu))},
    \end{split}
\ee
as already reported in $(\ref{eq:part_fct_sl})$ after imposing the restrictions (\ref{eq:scenario}).

We now turn to the integrals over $\hat{\zeta}$ and $x$, where the conditions (\ref{eq:scenario}) are necessary to avoid cumbersome expressions. The portion of the action (\ref{eq:action_torus_2}) relevant for carrying out the $\hat{\zeta}$-integral is 
\begin{equation}
     \begin{split}
        I_{\hat{\zeta}} = \frac{1}{2{\pi}} \int 2 \, d^2\sigma \left( \left( k -2 \right)  \p_z \hat{\zeta} \p_{\bar{z}} \hat{\zeta} - i  \bar{P} \p_z x \p_{\bar{z}} \hat{\zeta}  \right).
    \end{split}
\end{equation}
By introducing the rescaled field
\be
   \tilde{\zeta} = \sqrt{k-2} \,  \hat{\zeta},
\ee
the path integral over $\tilde{\zeta}$ is brought into the canonical form discussed in Appendix \ref{sec:path_int_freefields} with background field
\be
    A_z = \frac{-i\bar{P} \p_{z} x }{\sqrt{k-2}}, \qquad  A_{\bar{z}}=0.
\ee
Since the zero mode is excluded from the $\tilde{\zeta}$-integral, its contribution to the worldsheet partition function is 
\be \label{eq:part_fct_zeta}
    \mathcal{Z}_{\hat{\zeta}}  = \sqrt{k-2} \, \frac{e^{-\pi\tau_2\sum'_{m,n} \left( 1 -\frac{2m \tau_2}{i (m \bar{\tau}-n)}\right) A_{z}^{m,n} A_{z}^{-m,-n}}}{(\tau_2)^{\frac{1}{2}}|\eta(\tau)|^2},
\ee
where the primed sum is over $(m,n) \neq (0,0)$. 

To proceed further, we Fourier expand $x$ as
\be \label{eq:expansion_x}
    x = x_{cl} + \tilde{x} = x_{cl} + \sum_{m,n}{} x^{m,n}e^{\frac{1}{2\tau_2}(m(\bar{z}\tau-z \bar{\tau})+n(z-\bar{z}))},
\ee
where the classical solution $x_{cl}$ is constrained by the periodicity condition $x \sim x+2\pi R$ and takes the form
\be
    x_{cl} = \frac{R}{2 i\tau_2} \left( \omega_1 ( \bar{z} \tau-z \bar{\tau}  ) + \omega_2 (z - \bar{z} ) \right), \qquad \omega_1, \, \omega_2 \in \mathbb{Z}.
\ee
With this decomposition, the gauge field can be written as the sum of a classical and a fluctuating part
\be
    \begin{split}
        A_{z} = & A_{cl} + \tilde{A} = \frac{i \bar{P}}{\sqrt{k-2}} \frac{R}{2i\tau_2} (\omega_1 \bar{\tau} - \omega_2) \\
        & + \frac{i \bar{P}}{\sqrt{k-2}} \sum_{m,n}{}^{'}  \left(\frac{m \bar{\tau} -n}{2 \tau_2} \right) x^{-m,-n} e^{\frac{1}{2\tau_2}(m(\bar{z}\tau-z \bar{\tau})+n(z-\bar{z}))}.
    \end{split}
\ee
Plugging the above expression into (\ref{eq:part_fct_zeta}) yields
\be
    \mathcal{Z}_{\hat{\zeta}}= \sqrt{k-2} \, \frac{e^{-\frac{\pi\tau_2 \bar{P}^2}{k-2} \sum'_{m,n} \left( 1 -\frac{2m \tau_2}{i (m \bar{\tau}-n)}\right) \left( \frac{m \bar{\tau}-n}{2 \tau_2} \right)^2 x^{m,n} x^{-m,-n}}}{( \tau_2)^{\frac{1}{2}}|\eta(\tau)|^2}.
\ee

Finally, the purely $x$-dependent part of the action (\ref{eq:action_torus_2}) is given by
\be
    \begin{split}
        I_{x} = & \frac{1}{2{\pi}} \int 2 \, d^2\sigma  \left( \p_z x \p_{\bar{z}} x + \frac{\bar{P} \bar{U}^1_{n,m}(\tau|\bar{\alpha}, \beta, 0)}{2 \tau_2} \p_{z} x \right) \\
        = &  \frac{1}{2{\pi}} \int 2\, d^2\sigma  \left[ \p_z x_{cl} \p_{\bar{z}} x_{cl} + \frac{2 i \pi^2 \bar{P} \bar{U}^1_{n,m}(\tau|\bar{\alpha}, \beta, 0) R }{\tau_2}(\omega_1 \bar{\tau}-\omega_2) \right] \\
        & + \sum_{m,n}{}^{'}4\pi \tau_2 \left| \frac{- m \bar{\tau}+n}{2 \tau_2} \right|^2 x^{m,n} x^{-m,-n},
    \end{split}
\ee
where the last equality follows from the Fourier expansion (\ref{eq:expansion_x}) and the reality condition $(x^{m,n})^* = x^{-m,-n}$. The path integral over $x$ yields
\be
    \begin{split}
        \mathcal{Z}_{\hat{\zeta},x} = & \int Dx \,  \mathcal{Z}_{\hat{\zeta}} \, e^{-I_{x}} \\ 
        = & (2\pi R) \,\frac{2 \sqrt{k-2}}{|\eta(\tau)|^2} \, \sum_{\omega_1, \omega_2 \in \mathbb{Z}} e^{-\frac{\pi}{\tau_2} \left( | R(\omega_1\bar{\tau} - \omega_2)|^2 + i \bar{P} \bar{U}^1_{n,m}(\tau|\bar{\alpha}, \beta, 0) R (\omega_1\bar{\tau} - \omega_2) \right)}  \\
        & \sqrt{\prod_{m,n}{}^{'}\frac{1}{  \left| \frac{ m \tau-n}{2 \tau_2} \right|^2+\frac{ \bar{P}^2}{4(k-2) }  \left( 1 - \frac{2m \tau_2}{i (m \bar{\tau}-n)}\right) \left( \frac{m \bar{\tau}-n}{2 \tau_2} \right)^2}}.
    \end{split}
\ee
The infinite product can be evaluated using the $\zeta$-function regularization scheme, leading to
\be
    \begin{split}
        & \prod_{m,n}{}^{'} \left( \left| \frac{ m \tau-n}{2 \tau_2} \right|^2+\frac{ \bar{P}^2}{4(k-2) }  \left( 1 - \frac{2m \tau_2}{i (m \bar{\tau}-n)}\right) \left( \frac{m \bar{\tau}-n}{2 \tau_2} \right)^2 \right)  \\
        = & \prod_{m,n}{}^{'} \left( \frac{(n - m \tau)(n-m \bar{\tau})}{(2 \tau_2)^2} + \frac{ \bar{P}^2}{4(k-2) }  \frac{m \bar{\tau} -n +i2 m \tau_2}{m \bar{\tau} - n} \left( \frac{m \bar{\tau}-n}{2 \tau_2} \right)^2 \right)  \\
        =& \frac{(2 \tau_2)^2}{ 1 + \frac{ \bar{P}^2}{4(k-2) }  } \prod_{m,n}{}^{'} (n - m \tau)(n-m \bar{\tau}) = \frac{(4 \, \pi \, \tau_2)^2}{ 1 + \frac{ \bar{P}^2}{4(k-2) }  } \, \eta^2 \, \bar{\eta}^2.
    \end{split}
\ee
As a result, the contribution of the $\hat{\zeta}$ and $x$ path integrals to the worldsheet partition function takes the final form 
\be
    \begin{split}
       \mathcal{Z}_{\zeta,x} = \, \frac{R}{\tau_2 \, |\eta(\tau)|^4} \, \sqrt{k-2 + \frac{ \bar{P}^2}{4}} \sum_{\omega_1, \omega_2 \in \mathbb{Z}} e^{-\frac{\pi}{\tau_2}  \left( | R(\omega_1\bar{\tau} - \omega_2)|^2 + i \bar{P} \bar{U}^1_{n,m}(\tau|\bar{\alpha}, \beta, 0) R (\omega_1\bar{\tau} - \omega_2) \right)},
    \end{split}
\ee
as already reported in (\ref{eq:part_fct_zetax}) in the main text.

\section{Modular invariance}\label{sec:modular_inv}

In this appendix, we provide the technical details underlying the proof of modular invariance of the $\frac{SL(2, \mathbb{R}) \times U(1)}{U(1)}$ WZW model partition function discussed in Section \ref{sec:partition_fct}. The starting point is equation (\ref{eq:ws_part_fct}), already displayed in the main text: 
\begin{equation} 
    \begin{split}
        &\mathcal{Z}_{\text{wzw}} \left( \beta, k, \bar{P}, R \right) 
        = 2 \pi \beta \sqrt{k-2} \sqrt{k-2 + \frac{ \bar{P}^2}{4}} \sum_{n,m \in \mathbb{Z}} \int_{\mathcal{F}} \frac{d^2 \tau}{\tau_2^{5/2}} \int d^2 \alpha  |\eta|^4 (q \bar{q})^{- \frac{c_{\text{int}}}{24}} \\
        & \sum_{h, \bar{h}} D(h, \bar{h}) q^h \bar{q}^{\bar{h}} \, (2 \pi R) \sum_{\omega_1, \omega_2 \in \mathbb{Z}} e^{-\frac{\pi}{\tau_2}  \left( | R(\omega_1\bar{\tau} - \omega_2)|^2 + i \bar{P} \bar{U}^1_{n,m}(\tau| \bar{\alpha}, \beta, 0) R (\omega_1\bar{\tau} - \omega_2) \right)} \\
        & \frac{e^{-\frac{2 \pi}{\tau_2} \text{Im} \left( U_{n,m}^{1}(\bar{\tau}| \alpha, \beta, 0) \right) \text{Im} \left( \bar{U}^{1}_{n,m}(\tau| \bar{\alpha}, \beta, 0) \right) - \frac{k \pi}{\tau_2} U^{1}_{n,m}(\bar{\tau}| \alpha, \beta, 0) \bar{U}_{n,m}^{1}(\tau| \bar{\alpha}, \beta, 0)}}{\vartheta_{11}(\bar{\tau}, U_{n,m}^{1}(\bar{\tau}|\alpha, \beta, 0)) \vartheta_{11}(\tau, \bar{U}^{1}_{n,m}(\tau| \bar{\alpha}, \beta, 0))}.
    \end{split}
\end{equation}
This expression is manifestly invariant under modular $T$-transformation, provided the sums over $m$, $n$, $\omega_1$ and $\omega_2$ are treated appropriately. The invariance under $S$-transformation is less trivial and requires a more careful analysis. Under $S: \tau \rightarrow - \frac{1}{\tau}$ we have
\be
    \begin{split}
        U_{n,m}^{1}(\bar{\tau}|\alpha,\beta, 0) \xrightarrow[]S{} \frac{U_{-m,n}(\bar{\tau}|  \beta, 0) + \alpha \bar{\tau}}{\bar{\tau}},
    \end{split}
\ee
which shows that, by suitably exchanging $\omega_1$ and $\omega_2$, the contribution from the $x$-circle is modular invariant. 

To simplify the following analysis, we introduce 
\begin{equation}
    u = \bar{U}^{1}_{n,m}(\tau| \bar{\alpha}', \beta, 0), \qquad \bar{u} = U^{1}_{n,m}(\bar{\tau}| \alpha', \beta, 0).
\end{equation}
In terms of these new variables, the relevant transformations take the form
\be
    \begin{split}
        e^{-\frac{2 \pi}{\tau_2} \text{Im} (u) \text{Im}(\bar{u})} & \xrightarrow[]{S} e^{-\frac{2 \pi |\tau|^2}{\tau_2} \text{Im} \left( \frac{u}{\tau}\right) \text{Im} \left( \frac{\bar{u}}{\bar{\tau}}\right)},\\
        \frac{1}{\vartheta_{11}\left(\bar{\tau},\bar{u} \right) \vartheta_{11} \left(\tau, u \right)} & \xrightarrow[]{S} \frac{e^{ -i\pi  \left(  \frac{u^2}{\tau} -  \frac{ \bar{u}^2}{\bar{\tau}} \right)}} {|\tau| \vartheta_{11}\left(\bar{\tau},\bar{u} \right) \vartheta_{11} \left(\tau, u \right)}.
    \end{split}
\ee
Here we have flipped $(n,m) \to (-m,n)$ since both $m$ and $n$ are summed over. Next, performing the change of variables $\alpha' = \alpha \bar{\tau}$, $\bar{\alpha}'= \bar{\alpha} \tau$ introduces a Jacobian factor of $1/|\tau|^2$ in the corresponding measure. Combining this with the transformation of the moduli space measure, the condition for modular invariance reduces to\footnote{We are using the identity valid for any two complex numbers $a$ and $b$:
\be
(\text{Im} \left( ab \right))^2 =  \text{Im} \left( a^2b \right)\text{Im}(b)+\text{Im}(a)^2|b|^2,
\ee
}
\be
    \begin{split}
        &\frac{2 \pi |\tau|^2}{\tau_2} \text{Im} \left( \frac{\bar{u}}{\bar{\tau}}\right) \text{Im} \left( \frac{u}{\tau} \right)+i\pi  \left(  \frac{u^2}{\tau} - \frac{ \bar{u}^2}{\bar{\tau}} \right)=\frac{2 \pi}{\tau_2} \text{Im} \left( \bar{u}\right) \text{Im} \left( u \right)\\
        & \implies \text{Im} \left( \bar{u}\tau\right) \text{Im} \left( u\bar{\tau} \right)-\frac{ \left(  u^2\bar{\tau} -  \bar{u}^2\tau \right)}{2i} \text{Im}(\tau)=|\tau|^2 \text{Im} \left( \bar{u}\right) \text{Im} \left( u \right).
    \end{split}
\ee
This is indeed satisfied since $u=\bar{u}^*$.

\section{Computation of the target-space free energy}\label{sec:euclidean}

In this appendix, we conclude the computation of the worldsheet partition function of the charged 2D black hole initiated in Appendix \ref{sec:path_int_ads3}. As will be evident in the following, this computation closely parallels that of the worldsheet partition function of strings propagating on thermal $AdS_3$ \cite{Maldacena:2001}. The starting point is the single-string contribution to the free energy $(\ref{eq:free_2d})$, which we reproduce here for convenience
\be \label{eq:free_2d_app} 
    \begin{split}
        f_{\text{2D}} (\beta) = & 4 C \sum_{h, \bar{h}} D(h,\bar{h}) \sum_{w=0}^\infty \sum_{\ell, \bar{\ell} =0}^\infty \sum_{N, \bar{N} \in \mathbb{Z}}  \int d^2 \alpha \int_{-\frac{1}{2}}^{\frac{1}{2}} d \tau_1 \int_{\frac{\beta}{2 \pi(w+a_1+1)}}^{\frac{\beta}{2 \pi (w+a_1)}} \frac{d \tau_2}{\tau_2^2} e^{4 \pi \tau_2 \left( 1-\frac{1}{4(k-2)} \right)} \\
        & e^{- \frac{\pi k}{\tau_2} |\alpha|^2 + \frac{ 2 \pi \alpha_2^2}{\tau_2} - \left( k-2 -\frac{\bar{P}^2}{4} \right)\frac{\beta^2}{4 \pi \tau_2} + \left( \left(k-2 -\frac{\bar{P}^2}{4} \right)\alpha_2 - \frac{i \alpha_1 \bar{P}^2}{4} \right) \frac{\beta}{\tau_2} - \frac{\pi \bar{\alpha}^2 \bar{P}^2}{4 \tau_2} } \\
       & e^{-\left( \beta -2 \pi i \bar{\alpha}\right) \left( \frac{1}{2}+w+\ell \right) - \left( \beta + 2 \pi i \alpha \right) \left( \frac{1}{2}+w+\bar{\ell} \right)} \\
       & e^{2 \pi i \tau_1 \left( N +h -\bar{N} - \bar{h} \right)} e^{-2 \pi \tau_2 \left( N + h + \bar{N} + \bar{h} -w(w+1) \right)}, \\
       C= &  \, \pi^2 \, \sqrt{k-2} \, \sqrt{k-2+\frac{\bar{P}^{2}}{4}}.
    \end{split}
\ee
We start by explicitly imposing the decomposition (\ref{eq:gauge}): $\alpha = a_1 \bar{\tau}-a_2$ with $a_1, \, a_2 \in (0,1)$. The change of measure $d^2 \alpha = \tau_2 \, da_1 da_2$ introduces an additional factor of $\tau_2$. Since the resulting integrals over $a_1$ and $a_2$ are not Gaussian, we replace all terms involving them with their Fourier transforms and introduce the corresponding conjugate momenta $c_1$ and $c_2$. After integration over $c_1$, $c_2$ and $a_2$ we note that the integration over $\tau_1$ simply enforces the level matching condition
\begin{equation}\label{eq:level_matching}
    \int_{-\frac{1}{2}}^{\frac{1}{2}} d \tau_1 e^{2 \pi i \tau_1 \left( N +h -\bar{N} - \bar{h} \right)} = \delta_{N +h, \bar{N} + \bar{h}}.
\end{equation}
Overall, we are left with the expression
\be 
    \begin{split}
        f_{\text{2D}} (\beta) =& 4 C \sum_{h, \bar{h}} D(h,\bar{h}) \sum_{w=0}^\infty \sum_{\ell, \bar{\ell} =0}^\infty \sum_{N, \bar{N} \in \mathbb{Z}}\int_0^1 d a_1 \int_{\frac{\beta}{2 \pi(w+a_1+1)}}^{\frac{\beta}{2 \pi (w+a_1)}} \frac{d \tau_2}{\tau_2} e^{ -a \tau_2 - \frac{\beta^2 \, b}{4 \pi \tau_2} - \beta \, c},
    \end{split}
\ee
with coefficients
\begin{equation} \label{eq:coeff}
    \begin{split}
        a = \, & 4 \pi \left( N + h -\frac{w(w+1)}{2} -1 + \frac{1}{4(k-2)} \right), \\
        b = \, &  k-2-\frac{\bar{P}^2}{4}, \\
        c = \, & 1+\ell+\bar{\ell}+2w.
    \end{split}
\end{equation}
The above expression closely resembles the result in \cite{Maldacena:2001}. Crucially, the effect of the gauging procedure is encoded in the lower power of $\tau_2$ in the denominator ($\tau_2$ instead of $\tau_2^{3/2}$), which indicates that we are performing a 2D computation. It is also manifested in the modified bounds of the $\tau_2$ integration. 

Using the Gaussian identity  
\begin{equation}\label{eq:identity_gauss}
    e^{-\frac{\beta^2 b}{4 \pi \tau_2}} = \frac{8 \pi}{i \beta} \left( \frac{\tau_2}{b}\right)^{\frac{3}{2}}\int d \xi \, \xi \,  e^{-\frac{4 \pi \tau_2}{b} \xi^2 + 2 i \beta \xi}, 
\end{equation}
integration over $\tau_2$ yields
\begin{equation}\label{eq:f}
    \begin{split}
        & f_{\text{2D}} (\beta) = 4 C \sum_{h, \bar{h}} D(h,\bar{h}) \sum_{w=0}^\infty \sum_{\ell, \bar{\ell} =0}^\infty \sum_{N, \bar{N} \in \mathbb{Z}} \int_0^1 d a_1 \\
        & \frac{8 \pi}{i \beta b^{3/2}} \int d \xi \, \xi \,\frac{e^{2 i \beta \xi - \beta c}}{\left(\frac{4 \pi \xi^2}{b} + a \right)^{3/2}} \left[ \Gamma\left( \frac{3}{2}, \frac{\beta \left(\frac{4 \pi \xi^2}{b} + a \right)}{2 \pi (w + a_1 +1)} \right) - \Gamma\left( \frac{3}{2}, \frac{\beta \left(\frac{4 \pi \xi^2}{b} + a \right)}{2 \pi (w + a_1)} \right) \right],
    \end{split}
\end{equation}
where $\Gamma(s,z)$ represents the incomplete Gamma function \cite{Gammafct}. Using the identity
\begin{equation}\label{eq:id_gamma}
    \Gamma(s+1, z) = z^s \, e^{-z} + s\, \Gamma(s,z),
\end{equation}
we can decompose the $\xi-$integral in the second line of (\ref{eq:f}) as
\begin{equation}\label{eq:int_dec}
    \begin{split}
        &\frac{2}{i  \sqrt{\beta b}} \int d \xi \, \xi \, \frac{e^{2i \beta \xi - \beta c}}{ \xi^2 + \frac{ab}{4 \pi}}  \left[ \frac{e^{-\frac{\beta \left( \frac{2 \xi^2}{b} + \frac{a}{2 \pi} \right)}{w + a_1 +1}}}{\sqrt{2\pi(w + a_1 +1)}} - \frac{e^{-\frac{\beta \left( \frac{2 \xi^2}{b} + \frac{a}{2 \pi} \right)}{w + a_1}}}{\sqrt{2\pi(w + a_1)}} \right] \\
        +  & \frac{1}{\sqrt{4 \pi} i \beta} \int d \xi \,\xi \,\frac{e^{2 i \beta \xi - \beta c}}{\left( \xi^2 + \frac{ab}{4\pi}\right)^{3/2}} \left[ \Gamma\left( \frac{1}{2}, \frac{\beta \left( \frac{2 \xi^2}{b} + \frac{a}{2\pi}\right)}{w + a_1 +1} \right) - \Gamma\left( \frac{1}{2}, \frac{\beta  \left(\frac{2 \xi^2}{b} + \frac{a}{2 \pi} \right)}{w + a_1} \right) \right].
    \end{split}
\end{equation} 
To proceed, we observe that the term in the first line already has the appropriate structure to isolate the contribution from short strings, whereas the term in the second line requires further manipulation. In particular, using the identity 
\begin{equation}
    \frac{\xi}{\left(\xi^2 + \frac{ab}{4\pi}\right)^{3/2}} = - \frac{d}{d \xi} \left( \frac{1}{\sqrt{\xi^2 + \frac{ab}{4\pi}}} \right),
\end{equation}
we can integrate by parts the second-line contribution in (\ref{eq:int_dec}). In doing so, we also use the derivative of the incomplete Gamma function \cite{Gammafct}
\begin{equation}
    \frac{\partial \Gamma(s,z)}{\partial z} = - z^{s-1} e^{-z}.
\end{equation}
This procedure yields two contributions: one cancels exactly against the term in the first line of (\ref{eq:int_dec}), while the other looks
\begin{equation}
    \begin{split}
        \frac{1}{\sqrt{\pi}} \int d \xi \frac{e^{2 i \beta \xi - \beta c}}{\sqrt{\xi^2 + \frac{a b}{4 \pi}}} \left[ \Gamma\left( \frac{1}{2}, \frac{\beta \left( \frac{2 \xi^2}{b} + \frac{a}{2\pi}\right)}{w + a_1 +1} \right) - \Gamma\left( \frac{1}{2}, \frac{\beta  \left(\frac{2 \xi^2}{b} + \frac{a}{2 \pi} \right)}{w + a_1} \right) \right].
    \end{split}
\end{equation}
The net effect of this manipulation is then to increase the power of $(\xi^2 + \frac{a b}{4 \pi})$ in the denominator while preserving the same Gamma-function structure of the original expression (\ref{eq:f}). Applying the identity (\ref{eq:id_gamma}) once more, we obtain
\begin{equation}\label{eq:first_order}
    \begin{split}
        & \sqrt{\frac{b}{2 \pi \beta}} \int d \xi \, \frac{e^{2 i \beta \xi - \beta c}}{\xi^2 + \frac{ab}{4 \pi}} \left[ \sqrt{w + a_1 +1} \, e^{-\frac{\beta \left( \frac{2 \xi^2}{b} + \frac{a}{2 \pi} \right)}{w + a_1 +1}}- \sqrt{w + a_1} \,  e^{-\frac{\beta \left( \frac{2 \xi^2}{b} + \frac{a}{2 \pi} \right)}{w + a_1}}\right] \\
        & - \frac{1}{2 \sqrt{\pi}} \int d \xi \, \frac{e^{2 i \beta \xi - \beta c}}{\sqrt{\xi^2 + \frac{ab}{4 \pi}}} \left[ \Gamma\left( - \frac{1}{2}, \frac{\beta \left( \frac{2 \xi^2}{b} + \frac{a}{2\pi}\right)}{w + a_1 +1} \right) - \Gamma\left( - \frac{1}{2}, \frac{\beta  \left(\frac{2 \xi^2}{b} + \frac{a}{2 \pi} \right)}{w + a_1} \right)  \right].
    \end{split}
\end{equation} 

By the standard contour-shifting argument we can extract the short-string contribution from the integral in the first line. In particular, we note that the exponent in the first term can be expressed as a complete square if we set $\xi= s + \frac{i}{2}  b (w +a_1+1)$. Shifting the contour of the $\xi$-integral from $\text{Im}(\xi)=0$ to $\text{Im}(\xi)= \frac{b}{2} (w +a_1+1)$ makes the integral over $s$ real. In this process the integration contour crosses the poles in the range $0<\text{Im}(\xi)< \frac{b}{2} (w +a_1+1)$ and the integrand function picks up the corresponding residues at 
\begin{equation}
    - \frac{\xi^2}{b} = \frac{a}{4 \pi} <  \frac{b}{4} (w + a_1+1)^2.
\end{equation}
An analogous result holds for the second term, where we shift the contour by $\xi = s + \frac{i}{2}  b (w +a_1)$ and we pick up the residues at 
\begin{equation}
    - \frac{\xi^2}{b} = \frac{a}{4 \pi} <  \frac{b}{4} (w + a_1)^2.
\end{equation}
Since these residues have opposite sign we are effectively considering only the poles in the strip $\frac{b}{2} (w+a_1)< \text{Im}(\xi) < \frac{b}{2} (w+a_1 +1)$ corresponding to the interval
\begin{equation}\label{eq:second_constr}
     \frac{b}{4} (w + a_1)^2 < \frac{a}{4 \pi} <  \frac{b}{4} (w + a_1+1)^2,
\end{equation}
whose residues are
\begin{equation}\label{eq:s_t}
    -  \frac{\sqrt{2\pi(w + a_1+1)}}{2 \sqrt{\frac{a}{4 \pi}}} \frac{1}{\sqrt{\beta}} e^{-\beta \left(c+2\sqrt{\frac{a b}{4 \pi}}\right)} .
\end{equation}
The above result represents the short-string contribution to the free energy of the terms in the first line of (\ref{eq:first_order}). The corresponding long-string contribution is obtained by actually shifting the contours as discussed above. It results
\begin{equation}\label{eq:long_string}
    \begin{split}
        + &\frac{\sqrt{2 \pi b(w+a_1+1)}}{ 2 \pi  \sqrt{\beta}} \int d s \, \frac{e^{- \beta c -\frac{\beta}{w + a_1 +1} \left( \frac{2 s^2}{b} + \frac{b}{2}(w+a_1+1)^2 + \frac{a}{2 \pi} \right)}}{\left(s + \frac{i}{2} b (w+a_1+1)\right)^2 + \frac{ab}{4 \pi}}  \\
        - & \frac{\sqrt{2 \pi b(w+a_1)}}{2 \pi \sqrt{\beta}} \int d s  \, \frac{e^{- \beta c -\frac{\beta}{w + a_1} \left( \frac{2 s^2}{b} + \frac{b}{2}(w+a_1)^2 + \frac{a}{2 \pi} \right)}}{\left(s +\frac{i}{2} b (w+a_1)\right)^2 + \frac{ab}{4 \pi}}.
    \end{split}
\end{equation}
A point regarding both the short- and long-string spectra should be emphasized. The derivation presented above holds provided that $b > 0$. In the limit $b \to 0$, control over this expansion is lost, as manifested by the exponential divergence of the $s$-integral in \eqref{eq:long_string}. In Section \ref{sec:free_energy}, we speculate that this divergence in the long-string free energy is intimately related to the black hole/string transition.

To proceed, the long-string contribution (\ref{eq:long_string}) should be combined with the leftover integral in the second line of (\ref{eq:first_order}), which makes a full treatment of the free energy technically intractable. However, as discussed explicitly below (\ref{eq:free_2d}), the long-string contributions are not relevant in the large-$\beta$ limit of $f_{\text{2D}} (\beta)$. We will therefore neglect the above result in the following analysis.

In principle, we could further apply the identity (\ref{eq:id_gamma}) to the term appearing in the second line of (\ref{eq:first_order}). This would produce a contribution analogous to the first line in (\ref{eq:first_order}), but with a double pole, together with an additional term that preserves the same incomplete Gamma-function structure as in the second line of (\ref{eq:first_order}). The first of these contributions would generate another term proportional to $1/\sqrt{\beta}$ (albeit with a different coefficient), along with a subleading correction of order $1/\beta^{3/2}$. This procedure could be iterated, yielding a sequence of subleading corrections. However, since our interest lies only in the leading-order temperature dependence, it suffices to consider the result (\ref{eq:s_t}) arising from the first line of (\ref{eq:first_order}).

Overall, the leading short-string contribution to the free energy reads
\begin{equation}
    \begin{split}
        f_{\text{2D}} (\beta)\Big|_{\text{short strings}} = & - 2 C \int_0^1 da_1 \sum{\vphantom{\sum}}' D'(h,\bar{h}, N, \bar{N}, \ell, \bar{\ell}, w; a_1) \, \frac{1}{\sqrt{\beta}} \, e^{-\beta \left( c + 2\sqrt{\frac{a b}{4 \pi}}\right)},
    \end{split}
\end{equation}
where the primed sum over $h$, $\bar{h}$, $N$, $\bar{N}$, $\ell$, $\bar{\ell}$, $w$ is constrained by (\ref{eq:level_matching}) and (\ref{eq:second_constr}) and we define the coefficient
\begin{equation}
    D'(h,\bar{h}, N, \bar{N}, \ell, \bar{\ell}, w; a_1) = \sqrt{\frac{2 \pi( w + a_1+1)}{\frac{a}{4 \pi}}} \, D(h,\bar{h}).
\end{equation}

Finally, once the single-string contribution to the free energy is known, the worldsheet partition function of the charged 2D black hole can be obtained from ($\ref{eq:free_en_target_sp}$) by systematically incorporating all multi-string contributions. This yields
\begin{equation}
    \begin{split}
        & \mathcal{Z}_{2D} \Big|_{\text{short strings}} = - \beta \sum_{m=1}^\infty \, f_{2D}(m \, \beta) \Big|_{\text{short strings}} \\
        = & 2 C \int_0^1 da_1 \sum{\vphantom{\sum}}' D'(h,\bar{h}, N, \bar{N}, \ell, \bar{\ell}, w; a_1) \, \sqrt{\beta} \, \text{Li}_{1/2}\left( e^{-\beta \left( c + 2 \sqrt{\frac{ab}{4 \pi}} \right)} \right),
    \end{split}
\end{equation}
where $\text{Li}_s(z)$ is the polylogarithmic function of order $s$ \cite{Polylog}. This is exactly the result reported in (\ref{eq:2d_part_funct}) upon plugging in (\ref{eq:coeff}).

\newpage

\bibliographystyle{JHEP}
\bibliography{cite}

@article{Giveon:2003,
  author        = {Giveon, Amit and Rabinovici, Eliezer and Sever, Amit},
  title         = {{Beyond the singularity of the 2Dcharged black hole}},
  eprint        = {hep-th/0305140},
  archiveprefix = {arXiv},
  doi           = {10.1088/1126-6708/2003/07/055},
  journal       = {JHEP},
  volume        = {07},
  pages         = {055},
  year          = {2003}
}

@article{Giveon:2005,
  author        = {Giveon, Amit and Sever, Amit},
  title         = {{Strings in a 2-d extremal black hole}},
  eprint        = {hep-th/0412294},
  archiveprefix = {arXiv},
  doi           = {10.1088/1126-6708/2005/02/065},
  journal       = {JHEP},
  volume        = {02},
  pages         = {065},
  year          = {2005}
}

@article{Maldacena:2001,
  author        = {Maldacena, Juan and Ooguri, Hirosi and Son, John},
  title         = {Strings in AdS3 and the SL(2,R) WZW model. II: Euclidean black hole},
  eprint        = {hep-th/0005183},
  archiveprefix = {arXiv},
  doi           = {10.1063/1.1377039},
  journal       = {J.Math.Phys.},
  volume        = {42},
  pages         = {2961–2977},
  year          = {2001}
}

@article{Chen:2013,
  author        = {Chen, Wei-Ming and Ho, Pei-Ming and Kao, Hsien-chung and Khoo, Fech Scen and Matsuo, Yutaka},
  title         = {Partition function of a chiral boson on a 2-torus from the Floreanini\textendash{}Jackiw Lagrangian},
  eprint        = {1307.2172 [hep-th]},
  archiveprefix = {arXiv},
  doi           = {10.1093/ptep/ptu021},
  journal       = {PTEP},
  volume        = {03},
  pages         = {033},
  year          = {2014}
}

@article{Hanany:2002,
  author        = {Hanany, Amihay and Prezas, Nikolaos and Troost, Jan},
  title         = {The Partition Function of the Two-Dimensional Black Hole Conformal Field Theory},
  eprint        = {hep-th/0202129},
  archiveprefix = {arXiv},
  doi           = {10.1088/1126-6708/2002/04/014},
  journal       = {JHEP},
  volume        = {04},
  pages         = {014},
  year          = {2002}
}

@article{Ray:1973,
  author        = {D. B. Ray and I. M. Singer},
  title         = {Analytic torsion for complex manifolds},
  doi           = {10.2307/1970909},
  journal       = {Annals Math.},
  volume        = {98},
  pages         = {154-177},
  year          = {1973}
}

@conference{Gawedzki:1991,
    author = {Gawedzki, K.},
    booktitle = {NATO Advanced Study Institute: Part of New symmetry principles in quantum field theory},
    title = {Non-Compact WZW Conformal Field Theories},
    eprint = {hep-th/9110076},
    pages = {0247-274},
    year = {1991}
}

@article{Gawedzki:1988,
  author        = {Gawedzki, K. and Krzysztof, A.},
  title         = {G/h Conformal Field Theory from Gauged WZW Model},
  eprint        = {},
  archiveprefix = {},
  doi           = {10.1016/0370-2693(88)91081-7},
  journal       = {Phys.Lett.B},
  volume        = {215},
  pages         = {119-123},
  year          = {1988}
}

@article{Gawedzki:1989,
  author        = {Gawedzki, K. and Krzysztof, A.},
  title         = {Coset construction from functional integrals},
  eprint        = {},
  archiveprefix = {},
  doi           = {10.1016/0550-3213(89)90015-1},
  journal       = {Nuclear Physics B},
  volume        = {320},
  pages         = {625-668},
  year          = {1989}
}

@article{Chen:2022,
  author        = {Chen, Yiming and Maldacena, Juan},
  title         = {String scale black holes at large D},
  eprint        = {2106.02169 [hep-th]},
  archiveprefix = {arXiv},
  doi           = {10.1007/JHEP01(2022)095},
  journal       = {JHEP},
  volume        = {01},
  pages         = {095},
  year          = {2022}
}

@article{Berkooz:2007,
  author        = {Berkooz, Micha and Komargodski, Zohar and Reichmann, Dori},
  title         = {Thermal AdS3, BTZ and competing winding modes condensation},
  eprint        = {0706.0610 [hep-th]},
  archiveprefix = {arXiv},
  doi           = {10.1088/1126-6708/2007/12/020},
  journal       = {JHEP},
  volume        = {12},
  pages         = {020},
  year          = {2007}
}

@article{Chen:2021,
  author        = {Chen, Yiming and Maldacena, Juan and Witten, Edward},
  title         = {On the black hole/string transition},
  eprint        = {2109.08563 [hep-th]},
  archiveprefix = {arXiv},
  doi           = {10.1007/JHEP01(2023)103},
  journal       = {JHEP},
  volume        = {01},
  pages         = {103},
  year          = {2023}
}

@article{Giveon:2004,
  author        = {Giveon, Amit and Konechny, Anatoly and Rabinovici, Eliezer and Sever, Amit},
  title         = {On Thermodynamical Properties of Some Coset CFT Backgrounds},
  eprint        = {hep-th/0406131},
  archiveprefix = {arXiv},
  doi           = {10.1088/1126-6708/2004/07/076},
  journal       = {JHEP},
  volume        = {07},
  pages         = {076},
  year          = {2004}
}

@article{Horne:1992,
  author        = {Horne, James H. and Horowitz, Gary T.},
  title         = {Exact Black String Solutions in Three Dimensions},
  eprint        = {hep-th/9108001},
  archiveprefix = {arXiv},
  doi           = {10.1016/0550-3213(92)90536-K},
  journal       = {Nucl.Phys.B},
  volume        = {368},
  pages         = {444-462},
  year          = {1992}
}

@conference{Horowitz:1992,
    author = {Gary T. Horowitz},
    booktitle = {Trieste Spring School and Workshop: String Theory and Quantum Gravity},
    title = {The Dark Side of String Theory: Black Holes and Black Strings},
    eprint = {hep-th/9110076},
    archiveprefix = {arXiv},
    year = {1992}
}

@article{Atick:1988,
  author        = {Atick, Joseph J. and Witten, Edward},
  title         = {The Hagedorn Transition and the Number of Degrees of Freedom of String Theory},
  eprint        = {},
  archiveprefix = {},
  doi           = {10.1016/0550-3213(88)90151-4},
  journal       = {Nucl.Phys.B},
  volume        = {310},
  pages         = {291-334},
  year          = {1988}
}

@article{Ishibashi:1991,
  author        = {Ishibashi, N. and Li, M. and Steif, A.R.},
  title         = {Two-dimensional charged black holes in string theory},
  eprint        = {},
  archiveprefix = {},
  doi           = {10.1103/PhysRevLett.67.3336},
  journal       = {Phys.Rev.Lett.},
  volume        = {67},
  pages         = {3336-3338},
  year          = {1991}
}

@article{Witten:1991,
  author        = {Witten, Edward},
  title         = {On string theory and black holes},
  eprint        = {},
  archiveprefix = {},
  doi           = {10.1103/PhysRevD.44.314},
  journal       = {Phys.Rev.D},
  volume        = {44},
  pages         = {314-324},
  year          = {1991}
}

@article{Dijkgraaf:1992,
  author        = {Dijkgraaf, Robert and Verlinde, Herman L. and Verlinde, Erik P.},
  title         = {String propagation in a black hole geometry},
  eprint        = {},
  archiveprefix = {},
  doi           = {10.1016/0550-3213(92)90237-6},
  journal       = {Nucl.Phys.B},
  volume        = {371},
  pages         = {269-314},
  year          = {1992}
}

@article{Mandal:1991,
  author        = {Mandal, Gautam and Sengupta, Anirvan M. and Wadia, Spenta R.},
  title         = {Classical solutions of two-dimensional string theory},
  eprint        = {},
  archiveprefix = {},
  doi           = {10.1142/S0217732391001822},
  journal       = {Mod.Phys.Lett.A},
  volume        = {6},
  pages         = {269-314},
  year          = {1991}
}

@article{McGuigan:1992,
  author        = {McGuigan, Michael D. and Nappi, Chiara R. and Yost, Scott A.},
  title         = {Charged black holes in two-dimensional string theory},
  eprint        = {hep-th/9111038},
  archiveprefix = {arXiv},
  doi           = {10.1016/0550-3213(92)90039-E},
  journal       = {Nucl.Phys.B},
  volume        = {375},
  pages         = {421-450},
  year          = {1992}
}

@article{HorowitzStrominger:1991,
  author        = {Horowitz, Gary T. and Strominger, Andrew},
  title         = {Black strings and P-branes},
  eprint        = {},
  archiveprefix = {},
  doi           = {10.1016/0550-3213(91)90440-9},
  journal       = {Nucl.Phys.B},
  volume        = {360},
  pages         = {197-209},
  year          = {1991}
}

@article{Antoniadis:1994,
  author        = {Antoniadis, Ignatios and Ferrara, Sergio and Kounnas, Costas},
  title         = {Exact supersymmetric string solutions in curved gravitational backgrounds},
  eprint        = {hep-th/9402073},
  archiveprefix = {arXiv},
  doi           = {10.1016/0550-3213(94)90331-X},
  journal       = {Nucl.Phys.B},
  volume        = {421},
  pages         = {343-372},
  year          = {1994}
}

@article{Maldacena1:2001,
  author        = {Maldacena, Juan Martin and Ooguri, Hirosi},
  title         = {Strings in AdS(3) and SL(2,R) WZW model 1.: The Spectrum},
  eprint        = {hep-th/0001053},
  archiveprefix = {arXiv},
  doi           = {10.1063/1.1377273},
  journal       = {J.Math.Phys.},
  volume        = {42},
  pages         = {2929-2960},
  year          = {2001}
}

@article{Stanford:2017,
  author        = {Stanford, Douglas and Witten, Edward},
  title         = {Fermionic Localization of the Schwarzian Theory},
  eprint        = {1703.04612 [hep-th]},
  archiveprefix = {arXiv},
  doi           = {10.1007/JHEP10(2017)008},
  journal       = {JHEP},
  volume        = {10},
  pages         = {008},
  year          = {2017}
}

@article{Johnson:1994,
  author        = {Johnson, Clifford V.},
  title         = {Exact models of extremal dyonic 4-D black hole solutions of heterotic string theory},
  eprint        = {hep-th/9403192},
  archiveprefix = {arXiv},
  doi           = {10.1103/PhysRevD.50.4032},
  journal       = {Phys.Rev.D},
  volume        = {50},
  pages         = {4032-4050},
  year          = {1994}
}

@article{Tseytlin:1994,
  author        = {Tseytlin, Arkady A.},
  title         = {Conformal sigma models corresponding to gauged Wess-Zumino-Witten theories},
  eprint        = {hep-th/9302083},
  archiveprefix = {arXiv},
  doi           = {10.1016/0550-3213(94)90461-8},
  journal       = {Nucl.Phys.B},
  volume        = {411},
  pages         = {509-558},
  year          = {1994}
}

@article{Iliesiu:2021,
  author        = {Iliesiu, Luca V. and Turiaci, Gustavo J.},
  title         = {The statistical mechanics of near-extremal black holes},
  eprint        = {2003.02860 [hep-th]},
  archiveprefix = {arXiv},
  doi           = {10.1007/JHEP05(2021)145},
  journal       = {JHEP},
  volume        = {05},
  pages         = {145},
  year          = {2021}
}

@article{Rakic:2024,
  author        = {Rakic, Ilija and Rangamani, Mukund and Turiaci, Gustavo J.},
  title         = {Thermodynamics of the near-extremal Kerr spacetime},
  eprint        = {2310.04532 [hep-th]},
  archiveprefix = {arXiv},
  doi           = {10.1007/JHEP06(2024)011},
  journal       = {JHEP},
  volume        = {06},
  pages         = {011},
  year          = {2024}
}

@article{Kapec:2024,
  author        = {Kapec, Daniel and Sheta, Ahmed and Strominger, Andrew and Toldo, Chiara},
  title         = {Logarithmic Corrections to Kerr Thermodynamics},
  eprint        = {2310.00848 [hep-th]},
  archiveprefix = {arXiv},
  doi           = {10.1103/PhysRevLett.133.021601},
  journal       = {Phys.Rev.Lett.},
  volume        = {113},
  pages         = {},
  year          = {2024}
}

@article{Nappi:1992,
  author        = {Nappi, Chiara R. and Pasquinucci, Andrea},
  title         = {Thermodynamics of two-dimensional black holes},
  eprint        = {gr-qc/9208002},
  archiveprefix = {arXiv},
  doi           = {10.1142/S021773239200272X},
  journal       = {Mod.Phys.Lett.A},
  volume        = {7},
  pages         = {3337-3346},
  year          = {1992}
}

@article{Giveon11:1993,
  author        = {Giveon, Amit and Rabinovici, Eliezer and Tseytlin, Arkady A.},
  title         = {Heterotic string solutions and coset conformal field theories},
  eprint        = {hep-th/9304155},
  archiveprefix = {arXiv},
  doi           = {10.1016/0550-3213(93)90583-B},
  journal       = {Nucl.Phys.B},
  volume        = {049},
  pages         = {339-362},
  year          = {1993}
}

@article{Lowe:1994,
  author        = {Lowe, David A. and Strominger, Andrew},
  title         = {Exact four-dimensional dyonic black holes and Bertotti-Robinson space-times in string theory},
  eprint        = {hep-th/9403186},
  archiveprefix = {arXiv},
  doi           = {10.1103/PhysRevLett.73.1468},
  journal       = {Phys.Rev.Lett.},
  volume        = {73},
  pages         = {1468-1471},
  year          = {1994}
}

@article{Strominger:1998,
  author        = {Strominger, Andrew},
  title         = {AdS(2) quantum gravity and string theory},
  eprint        = {hep-th/9809027},
  archiveprefix = {arXiv},
  doi           = {10.1088/1126-6708/1999/01/007},
  journal       = {JHEP},
  volume        = {01},
  pages         = {007},
  year          = {1998}
}

@article{Ferko:2025,
  author        = {Ferko, Christian and Murthy, Sameer and Rangamani, Mukund},
  title         = {Strings in $AdS_3$: one-loop partition function and near-extremal BTZ thermodynamics},
  eprint        = {2408.14567 [hep-th]},
  archiveprefix = {arXiv},
  doi           = {10.1007/JHEP05(2025)010},
  journal       = {JHEP},
  volume        = {05},
  pages         = {010},
  year          = {2025}
}

@article{Eguchi:2011,
  author        = {Eguchi, Tohru and Sugawara, Yuji},
  title         = {Non-holomorphic Modular Forms and SL(2,R)/U(1) Superconformal Field Theory},
  eprint        = {1012.5721 [hep-th]},
  archiveprefix = {arXiv},
  doi           = {10.1007/JHEP03(2011)107},
  journal       = {JHEP},
  volume        = {03},
  pages         = {107},
  year          = {2011}
}

@article{Giveon:1994,
  author        = {Giveon, Amit and Porrati, Massimo and Rabinovici, Eliezer},
  title         = {Target space duality in string theory},
  eprint        = {hep-th/9401139},
  archiveprefix = {arXiv},
  doi           = {10.1016/0370-1573(94)90070-1},
  journal       = {Phys.Rept.},
  volume        = {244},
  pages         = {77-202},
  year          = {1994}
}

@article{Horowitz:1993,
  author        = {Horowitz, Gary T. and Welch, Dean L.},
  title         = {Exact three-dimensional black holes in string theory},
  eprint        = {hep-th/9302126},
  archiveprefix = {arXiv},
  doi           = {10.1103/PhysRevLett.71.328},
  journal       = {Phys.Rev.Lett.},
  volume        = {71},
  pages         = {328-331},
  year          = {1993}
}

@article{Gibbons:1992,
  author        = {Gibbons, G.W. and Perry, M.J.},
  title         = {The Physics of 2-D stringy space-times},
  eprint        = {hep-th/9204090},
  archiveprefix = {arXiv},
  doi           = {10.1142/S0218271892000161},
  journal       = {Int.J.Mod.Phys.D},
  volume        = {1},
  pages         = {335-354},
  year          = {1992}
}

@article{Banerjee:2011,
  author        = {Banerjee, Shamik and Gupta, Rajesh Kumar and Sen, Ashoke},
  title         = {Logarithmic Corrections to Extremal Black Hole Entropy from Quantum Entropy Function},
  eprint        = {1005.3044 [hep-th]},
  archiveprefix = {arXiv},
  doi           = {10.1007/JHEP03(2011)147},
  journal       = {JHEP},
  volume        = {03},
  pages         = {147},
  year          = {2011}
}

@article{Banerjee1:2011,
  author        = {Banerjee, Shamik and Gupta, Rajesh Kumar and Mandal, Ipsita and Sen, Ashoke},
  title         = {Logarithmic Corrections to N=4 and N=8 Black Hole Entropy: A One Loop Test of Quantum Gravity},
  eprint        = {1106.0080 [hep-th]},
  archiveprefix = {arXiv},
  doi           = {10.1007/JHEP11(2011)143},
  journal       = {JHEP},
  volume        = {11},
  pages         = {143},
  year          = {2011}
}

@article{Sen:2012,
  author        = {Sen, Ashoke},
  title         = {Logarithmic Corrections to Rotating Extremal Black Hole Entropy in Four and Five Dimensions},
  eprint        = {1109.3706 [hep-th]},
  archiveprefix = {arXiv},
  doi           = {10.1007/s10714-012-1373-0},
  journal       = {Gen.Rel.Grav.},
  volume        = {44},
  pages         = {1947-1991},
  year          = {2012}
}

@article{Ghosh:2020,
  author        = {Ghosh, Animik and Maxfield, Henry and Turiaci, Gustavo J.},
  title         = {A universal Schwarzian sector in two-dimensional conformal field theories},
  eprint        = {1912.07654 [hep-th]},
  archiveprefix = {arXiv},
  doi           = {10.1007/JHEP05(2020)104},
  journal       = {JHEP},
  volume        = {05},
  pages         = {104},
  year          = {2020}
}

@article{Iliesiu:2020,
  author        = {Iliesiu, Luca V. and Kruthoff, Jorrit and Turiaci, Gustavo J. and Verlinde, Herman},
  title         = {JT gravity at finite cutoff},
  eprint        = {2004.07242 [hep-th]},
  archiveprefix = {arXiv},
  doi           = {10.21468/SciPostPhys.9.2.023},
  journal       = {SciPost Phys.},
  volume        = {9},
  pages         = {023},
  year          = {2020}
}

@article{Heydeman:2022,
  author        = {Heydeman, Matthew and Iliesiu, Luca V. and Turiaci, Gustavo J. and Zhao, Wenli},
  title         = {The statistical mechanics of near-BPS black holes},
  eprint        = {2011.01953 [hep-th]},
  archiveprefix = {arXiv},
  doi           = {10.1088/1751-8121/ac3be9},
  journal       = {J.Phys.A},
  volume        = {55},
  pages         = {},
  year          = {2022}
}

@article{Nayak:2018,
  author        = {Nayak, Pranjal and Shukla, Ashish and Soni, Ronak M. and Trivedi, Sandip P. and Vishal, V.},
  title         = {On the Dynamics of Near-Extremal Black Holes},
  eprint        = {1802.09547 [hep-th]},
  archiveprefix = {arXiv},
  doi           = {10.1007/JHEP09(2018)048},
  journal       = {JHEP},
  volume        = {09},
  pages         = {048},
  year          = {2018}
}

@article{Iliesiu:2022,
  author        = {Iliesiu, Luca V. and Murthy, Sameer and Turiaci, Gustavo J.},
  title         = {Revisiting the logarithmic corrections to the black hole entropy},
  eprint        = {2209.13608 [hep-th]},
  archiveprefix = {arXiv},
  doi           = {10.1007/JHEP07(2025)058},
  journal       = {JHEP},
  volume        = {07},
  pages         = {058},
  year          = {2022}
}

@article{Mertens:2023,
  author        = {Mertens, Thomas G. and Turiaci, Gustavo J.},
  title         = {Solvable models of quantum black holes: a review on Jackiw–Teitelboim gravity},
  eprint        = {2210.10846 [hep-th]},
  archiveprefix = {arXiv},
  doi           = {10.1007/s41114-023-00046-1},
  journal       = {Living Rev.Rel.},
  volume        = {26},
  pages         = {1},
  year          = {2023}
}

@article{Iliesiu:2025,
  author        = {Iliesiu, Luca V. and Murthy, Sameer and Turiaci, Gustavo J.},
  title         = {Black hole microstate counting from the gravitational path integral},
  eprint        = {2209.13602 [hep-th]},
  archiveprefix = {arXiv},
  doi           = {10.1007/JHEP08(2025)152},
  journal       = {JHEP},
  volume        = {08},
  pages         = {152},
  year          = {2025}
}

@conference{Dunne:1998,
    author = {Dunne, Gerald V.},
    booktitle = {Les Houches Summer School of Theoretical Physics: Part of Topological Aspects of Low-dimensional Systems},
    title = {Aspects of Chern-Simons theory},
    eprint = {hep-th/9902115},
    archiveprefix = {arXiv},
    year = {1998}
}

@article{Kazakov:2002,
  author        = {Kazakov, Vladimir and Kostov, Ivan K. and Kutasov, David},
  title         = {A Matrix model for the two-dimensional black hole},
  eprint        = {hep-th/0101011},
  archiveprefix = {arXiv},
  doi           = {10.1016/S0550-3213(01)00606-X},
  journal       = {Nucl.Phys.B},
  volume        = {622},
  pages         = {141-188},
  year          = {2002}
}

@article{Hikida:2009,
  author        = {Hikida, Yasuaki and Schomerus, Volker},
  title         = {The FZZ-Duality Conjecture: A Proof},
  eprint        = {0805.3931 [hep-th]},
  archiveprefix = {arXiv},
  doi           = {10.1088/1126-6708/2009/03/095},
  journal       = {JHEP},
  volume        = {03},
  pages         = {095},
  year          = {2009}
}

@article{Teschner:1999,
  author        = {Teschner, J.},
  title         = {On structure constants and fusion rules in the SL(2,C) / SU(2) WZNW model},
  eprint        = {hep-th/9712256},
  archiveprefix = {arXiv},
  doi           = {10.1016/S0550-3213(99)00072-3},
  journal       = {Nucl.Phys.B},
  volume        = {546},
  pages         = {390-422},
  year          = {1999}
}

@article{Teschner:2000,
  author        = {Teschner, J.},
  title         = {Operator product expansion and factorization in the H+(3) WZNW model},
  eprint        = {hep-th/9906215},
  archiveprefix = {arXiv},
  doi           = {10.1016/S0550-3213(99)00785-3},
  journal       = {Nucl.Phys.B},
  volume        = {571},
  pages         = {555-582},
  year          = {2000}
}

@article{Ribault:2005,
  author        = {Ribault, Sylvain and Teschner, Joerg},
  title         = {H+(3)-WZNW correlators from Liouville theory},
  eprint        = {hep-th/0502048},
  archiveprefix = {arXiv},
  doi           = {10.1088/1126-6708/2005/06/014},
  journal       = {JHEP},
  volume        = {06},
  pages         = {014},
  year          = {2005}
}

@article{Fukuda:2001,
  author        = {Fukuda, Takeshi and Hosomichi, Kazuo},
  title         = {Three point functions in sine-Liouville theory},
  eprint        = {hep-th/0105217},
  archiveprefix = {arXiv},
  doi           = {10.1088/1126-6708/2001/09/003},
  journal       = {JHEP},
  volume        = {09},
  pages         = {003},
  year          = {2001}
}

@article{Giribet:2021,
  author        = {Giribet, Gaston},
  title         = {String correlators in AdS3 from FZZ duality},
  eprint        = {2110.04197 [hep-th]},
  archiveprefix = {arXiv},
  doi           = {10.1007/JHEP12(2021)012},
  journal       = {JHEP},
  volume        = {12},
  pages         = {012},
  year          = {2021}
}

@article{Horowitz:1998,
  author        = {Horowitz, Gary T. and Polchinski, Joseph},
  title         = {Selfgravitating fundamental strings},
  eprint        = {hep-th/9707170},
  archiveprefix = {arXiv},
  doi           = {10.1103/PhysRevD.57.2557},
  journal       = {Phys.Rev.D},
  volume        = {57},
  pages         = {2557-2563},
  year          = {1998}
}

@article{Giveon:2007,
  author        = {Giveon, Amit and Kutasov, David},
  title         = {Fundamental strings and black holes},
  eprint        = {hep-th/0611062},
  archiveprefix = {arXiv},
  doi           = {10.1088/1126-6708/2007/01/071},
  journal       = {JHEP},
  volume        = {01},
  pages         = {071},
  year          = {2007}
}

@article{Kutasov:2005,
  author        = {Kutasov, David},
  title         = {Accelerating branes and the string/black hole transition},
  eprint        = {hep-th/0509170},
  archiveprefix = {arXiv},
  doi           = {},
  journal       = {},
  volume        = {},
  pages         = {},
  year          = {2005}
}

@article{Brustein:2021,
  author        = {Brustein, Ram and Zigdon, Yoav},
  title         = {Black hole entropy sourced by string winding condensate},
  eprint        = {2107.09001 [hep-th]},
  archiveprefix = {arXiv},
  doi           = {10.1007/JHEP10(2021)219},
  journal       = {JHEP},
  volume        = {10},
  pages         = {219},
  year          = {2021}
}

@misc{Polylog,
  author       = {{National Institute of Standards and Technology}},
  title        = {NIST Digital Library of Mathematical Functions, Chapter 25.12: Polylogarithms},
  year         = {Release 1.2.6 of 2026-03-15},
  url          = {https://dlmf.nist.gov/25.12}
}

@article{Kolanowski:2025,
  author        = {Kolanowski, Maciej and Marolf, Donald and Rakic, Ilija and Rangamani, Mukund and Turiaci, Gustavo J.},
  title         = {Looking at extremal black holes from very far away},
  eprint        = {2409.16248 [hep-th]},
  archiveprefix = {arXiv},
  doi           = {10.1007/JHEP04(2025)020},
  journal       = {JHEP},
  volume        = {04},
  pages         = {020},
  year          = {2025}
}

@article{Sachdev:2019,
  author        = {Sachdev, Subir},
  title         = {Universal low temperature theory of charged black holes with AdS2 horizons},
  eprint        = {1902.04078 [hep-th]},
  archiveprefix = {arXiv},
  doi           = {10.1063/1.5092726},
  journal       = {J.Math.Phys.},
  volume        = {60},
  pages         = {5},
  year          = {2019}
}

@book{Polchinski:2007,
    author        = {Polchinski, J.},
    title         = {String theory. Vol. 1: An introduction to the bosonic string},
    publisher     = {Cambridge
Monographs on Mathematical Physics, Cambridge University Press},
    year          = {2007}
}

@misc{Gammafct,
  author       = {{National Institute of Standards and Technology}},
  title        = {NIST Digital Library of Mathematical Functions, Chapter 8: Incomplete Gamma and Related Functions},
  year         = {Release 1.2.6 of 2026-03-15},
  url          = {https://dlmf.nist.gov/8}
}

@article{Sfetsos:1993,
  author        = {Sfetsos, Konstadinos},
  title         = {Conformally exact results for SL(2,R) x SO(1,1)(d-2) / SO(1,1) coset models},
  eprint        = {hep-th/9206048},
  archiveprefix = {arXiv},
  doi           = {10.1016/0550-3213(93)90327-L},
  journal       = {Nucl.Phys.B},
  volume        = {389},
  pages         = {424-444},
  year          = {1993}
}

@article{Bars:1993,
  author        = {Bars, I. and Sfetsos, K.},
  title         = {Exact effective action and space-time geometry n gauged WZW models},
  eprint        = {hep-th/9301047},
  archiveprefix = {arXiv},
  doi           = {10.1103/PhysRevD.48.844},
  journal       = {Phys.Rev.D},
  volume        = {48},
  pages         = {844-852},
  year          = {1993}
}

@article{Maulik:2026,
  author        = {Maulik, Sabyasachi and Zayas, Leopoldo A. Pando and Ray, Augniva and Zhang, Jingchao},
  title         = {Universality in logarithmic temperature corrections to near-extremal rotating black hole thermodynamics in various dimensions},
  eprint        = {2401.16507 [hep-th]},
  archiveprefix = {arXiv},
  doi           = {10.1007/JHEP01(2026)156},
  journal       = {JHEP},
  volume        = {01},
  pages         = {034},
  year          = {2026}
}

@article{Maulik2:2026,
  author        = {Maulik, Sabyasachi and Mitra, Arpita and Mukherjee, Debangshu and Ray, Augniva},
  title         = {Logarithmic corrections to near-extremal entropy of charged de Sitter black holes},
  eprint        = {2503.08617 [hep-th]},
  archiveprefix = {arXiv},
  doi           = {10.1007/JHEP01(2026)156},
  journal       = {JHEP},
  volume        = {01},
  pages         = {156},
  year          = {2026}
}

@article{Giveon_transition:2005,
  author        = {Giveon, A. and Kutasov, D. and Rabinovici, E. and Sever, A.},
  title         = {Phases of quantum gravity in AdS(3) and linear dilaton backgrounds},
  eprint        = {hep-th/0503121},
  archiveprefix = {arXiv},
  doi           = {10.1016/j.nuclphysb.2005.04.015},
  journal       = {Nucl.Phys.B},
  volume        = {719},
  pages         = {3-34},
  year          = {2005}
}

@article{Giveon:2006,
  author        = {Giveon, Amit and Kutasov, David},
  title         = {The Charged black hole/string transition},
  eprint        = {hep-th/0510211},
  archiveprefix = {arXiv},
  doi           = {10.1088/1126-6708/2006/01/120},
  journal       = {JHEP},
  volume        = {01},
  pages         = {120},
  year          = {2006}
}

@article{Horowitz:1997,
  author        = {Horowitz, Gary T. and Polchinski, Joseph},
  title         = {A Correspondence principle for black holes and strings},
  eprint        = {hep-th/9612146},
  archiveprefix = {arXiv},
  doi           ={10.1103/PhysRevD.55.6189},
  journal       = {Phys.Rev.D},
  volume        = {55},
  pages         = {6189-6197},
  year          = {1997}
}

\end{document}